\documentclass[12pt]{article}
\usepackage{color}
\usepackage{latexsym}
\usepackage{epsfig,amssymb,euscript, mathrsfs,cite}
\usepackage{amsmath}
\definecolor{MyDarkBlue}{rgb}{0.15,0.15,0.45}
\usepackage[linktocpage=true]{hyperref}
\usepackage{enumerate}
\hypersetup{
colorlinks=true,
citecolor=MyDarkBlue,
linkcolor=MyDarkBlue,
urlcolor=MyDarkBlue,
pdfauthor={},
pdftitle={},
pdfsubject={hep-th}
}
\newsavebox{\ns}
\newsavebox{\dbrane}
\newsavebox{\dbshort}

\def\be{\begin{equation}}
\def\ee{\end{equation}}
\def\bea{\begin{eqnarray}}
\def\eea{\end{eqnarray}}

\newcommand{\nn}{\nonumber}

\newcommand\Z{\mathbb{Z}}

\newcommand\C{\mathbb{C}}

\newcommand\diff{\mathrm{d}}

\newcommand{\dd}{\mathrm{d}}

\newcommand{\ii}{\mathrm{i}}

\newcommand{\ex}{\mathrm{e}}
\newcommand{\vol}{\mathrm{vol}}

\newcommand{\mt}{\mathrm{T}}

\newlength{\sswidth}

\newcommand{\hook}{\mathbin{\rule[.2ex]{.4em}{.03em}\rule[.2ex]{.03em}{.9ex}}}
\newcommand{\alphaangle}{\vartheta}

\newcommand{\Spin}{\mathcal{S}}
\newcommand{\rSUSY}{\rho}
\newcommand{\Sf}{S^{(0)}}
\newcommand{\alphaanglef}{\alphaangle^{(0)}}
\newcommand{\Kf}{K^{(0)}}
\newcommand{\Jf}{J^{(0)}}
\newcommand{\Omegaf}{\Omega^{(0)}}
\newcommand{\reeb}{\xi}

\numberwithin{equation}{section}       

\begin{document}

\bibliographystyle{utphys}

\begin{titlepage}

\begin{center}

\today

\vskip 2.3 cm 

\vskip 5mm

{\Large \bf Supersymmetric solutions to Euclidean }

\vskip 5mm

{\Large \bf Romans supergravity}

\vskip 15mm

{Luis F. Alday, Martin Fluder, Carolina Matte Gregory,}\\ \vskip 3mm
{ Paul Richmond and James Sparks}

\vspace{1cm}
\centerline{{\it Mathematical Institute, University of Oxford,}}
\centerline{{\it Andrew Wiles Building, Radcliffe Observatory Quarter,}}
\centerline{{\it Woodstock Road, Oxford, OX2 6GG, UK}}

\end{center}

\vskip 2 cm

\begin{abstract}
\noindent We study Euclidean Romans supergravity in six dimensions 
with a non-trivial Abelian R-symmetry gauge field. We show that 
supersymmetric solutions are in one-to-one correspondence with solutions to
a set of differential constraints on an $SU(2)$ structure. As an application 
of our results we (i) show that this structure reduces 
at a conformal boundary to the five-dimensional rigid supersymmetric geometry previously studied by the authors, 
(ii) find a general expression 
for the holographic dual of the VEV of a BPS Wilson loop, matching an 
exact field theory computation, (iii)  construct holographic 
duals to squashed Sasaki-Einstein backgrounds, again matching to a
field theory computation, and (iv) find new analytic solutions. 

\end{abstract}

\end{titlepage}

\pagestyle{plain}
\setcounter{page}{1}
\newcounter{bean}
\baselineskip18pt
\tableofcontents


\section{Introduction}\label{SecIntroduction}

Advances in localization techniques applied to gauge theories have led to exact results for supersymmetric observables on general backgrounds. 
In three and four dimensions it turns out that such observables depend 
on only a small number of parameters of the full parameter space of the background \cite{Alday:2013lba, Closset:2013vra}.\footnote{For a different approach see also~\cite{Imbimbo:2014pla}.} 
Rigid  supersymmetric gauge theories  in five-dimensional curved backgrounds  have been constructed and studied in a series of papers \cite{Kallen:2012cs, Hosomichi:2012ek, Kallen:2012va, Kim:2012ava, Imamura:2012xg, Imamura:2012bm,  Qiu:2013pta, Qiu:2013aga, Schmude:2014lfa, Pan:2013uoa, Imamura:2014ima, Alday:2015lta, Pan:2014bwa, Pan:2015nba, Pini:2015xha}. 
In the approach of \cite{Alday:2015lta} these rigid backgrounds are equipped with a transversely holomorphic foliation.
Inspired by the lower-dimensional results of  \cite{Alday:2013lba, Closset:2013vra} it was conjectured 
that supersymmetric observables depend only on this foliation. 
In this paper we systematically study supersymmetric solutions to 
Euclidean Romans supergravity in six dimensions. Our aim is to compute  observables of interest for gauge/gravity duality, 
and in particular understand the conjecture of \cite{Alday:2015lta} from a holographic perspective.

Our starting point is to show that real Euclidean supersymmetric solutions to 
Romans $F(4)$ gauged supergravity, with 
a non-trivial Abelian R-symmetry gauge field, have a canonical $SU(2)$ structure determined by the Killing spinor. 
More precisely we show that supersymmetry together with the equations of motion
are equivalent to a set of differential constraints on this $SU(2)$ structure. 
This geometric formulation then leads to a number of interesting applications. 
First, we show that this  structure extends into the bulk the 
conformal boundary $SU(2)$ structure studied in \cite{Alday:2015lta}. 
This allows for the construction of gravity duals 
to families of five-dimensional gauge theories on rigid backgrounds.  
As another application we extend several of the results in \cite{Alday:2014rxa, Alday:2014bta}. 
In the latter we 
constructed supergravity solutions with squashed five-sphere boundaries, 
and computed the holographic free energy and certain BPS Wilson loops. 
In the present paper we extend these results to new families of solutions, 
in general with different topology. In particular this includes squashed 
Sasaki-Einstein conformal boundaries, together with new analytic 
solutions. Furthermore, in \cite{Alday:2014rxa, Alday:2014bta} 
we conjectured  a general  formula for the VEV of a BPS Wilson loop, 
both in field theory and in supergravity. In this paper 
the supergravity conjecture is proven.

The outline of the paper is as follows. Section \ref{SecConditions} 
contains a general analysis of Euclidean supersymmetric solutions to 
Romans supergravity, recasting the conditions in terms of a canonical local $SU(2)$ structure. 
In section 
\ref{SecApplications} we present a number of applications of our formalism.
 Our conclusions are presented in section \ref{SecDiscussion}. 
A number of technical details have been included in five appendices.

\section{Conditions for supersymmetry}\label{SecConditions}

\subsection{Euclidean Romans supergravity}

The bosonic fields of the six-dimensional Romans supergravity theory \cite{Romans:1985tw} consist of the metric, a scalar field $X=\exp(-\tfrac{\phi}{2\sqrt{2}})$ where $\phi$ is the dilaton, a two-form potential $B$, together with an $SO(3)_R\sim SU(2)_R$ R-symmetry gauge field $A^i$ with field strength $F^i=\diff A^i-\frac{1}{2}\varepsilon_{ijk}A^j\wedge A^k$, where $i=1,2,3$. 
Here we are working in a gauge in which the Stueckelberg one-form is zero, and we set 
the  gauge coupling constant to 1. 
  The Euclidean signature equations of motion are \cite{Alday:2014bta}
  \bea\label{FullEOM}
  \diff\left(X^{-1}*\dd X\right) &=& -  \left(\tfrac{1}{6}X^{-6}-\tfrac{2}{3}X^{-2}+\tfrac{1}{2}X^2\right)*1 \nn \\
&&-\tfrac{1}{8}X^{-2}\left(\tfrac{4}{9}B\wedge *B+F^i\wedge *F^i\right) + \tfrac{1}{4}X^4H\wedge *H ~,\nn\\
\diff\left(X^4 * H\right) &=& \tfrac{2\, \ii}{9}B\wedge B + \tfrac{\ii}{2}F^i\wedge F^i +\tfrac{4}{9}  X^{-2}*B~,\nn\\
D(X^{-2}*F^i) & = & - \ii F^i\wedge H~.
\eea
Here $H=\diff B$ and
$D\omega^i=\dd\omega^i - \varepsilon_{ijk}A^j\wedge \omega^k$ is the 
$SO(3)$ covariant derivative. Notice that the theory 
contains Chern-Simons-type couplings, that become purely imaginary in Euclidean signature.
The Einstein equation is
\bea\label{Einstein}
R_{\mu\nu} &=& 4X^{-2}\partial_\mu X\partial_\nu X + \left(\tfrac{1}{18}X^{-6}-\tfrac{2}{3}X^{-2}-\tfrac{1}{2}X^2\right) g_{\mu\nu} + \tfrac{1}{4}X^4\left(H^2_{\mu\nu}-\tfrac{1}{6}H^2g_{\mu\nu}\right) \nn\\
&& +  \tfrac{2}{9}X^{-2}\left(B^2_{\mu\nu}-\tfrac{1}{8}B^2g_{\mu\nu}\right) +  \tfrac{1}{2}X^{-2}\left((F^i)^2_{\mu\nu}-\tfrac{1}{8}(F^i)^2g_{\mu\nu}\right)~,
\eea
where $B^2_{\mu\nu} = B_{\mu\rho} B_\nu{}^\rho$, $H^2_{\mu\nu}=H_{\mu\rho\sigma}H_{\nu}^{\ \rho\sigma}$.
  
A  solution is 
 supersymmetric provided there 
exists a non-trivial $SU(2)_R$ doublet of Dirac spinors $\epsilon_I$, $I=1,2$, satisfying 
the following Killing spinor  and dilatino equations
\bea
D_\mu \epsilon_I & =&  \tfrac{\ii}{4\sqrt{2}}  ( X + \tfrac{1}{3} X^{-3} ) \Gamma_\mu \Gamma_7 \epsilon_I - \tfrac{\ii}{24\sqrt{2}} X^{-1} B_{\nu\rho} ( \Gamma_\mu{}^{\nu\rho} - 6 \delta_\mu{}^\nu \Gamma^\rho ) \epsilon_I \nn  \\
&& - \tfrac{1}{48} X^2 H_{\nu\rho\sigma} \Gamma^{\nu\rho\sigma} \Gamma_\mu \Gamma_7 \epsilon_I
+ \tfrac{1}{16\sqrt{2}}X^{-1} F_{\nu\rho}^i ( \Gamma_\mu{}^{\nu\rho} - 6 \delta_\mu{}^\nu \Gamma^\rho ) \Gamma_7 ( \sigma_i )_I{}^J \epsilon_J ~,\label{KSE}\\
0 & = & - \ii X^{-1} \partial_\mu X \Gamma^\mu \epsilon_ I + \tfrac{1}{2\sqrt{2}}  \left( X - X^{-3} \right) \Gamma_7 \epsilon_I + \tfrac{\ii}{24} X^2 H_{\mu\nu\rho} \Gamma^{\mu\nu\rho} \Gamma_7 \epsilon_I \nonumber \\
&&- \tfrac{1}{12\sqrt{2}} X^{-1} B_{\mu\nu} \Gamma^{\mu\nu} \epsilon_I - \tfrac{\ii}{8\sqrt{2}} X^{-1} F^i_{\mu\nu} \Gamma^{\mu\nu} \Gamma_7 ( \sigma_i )_I{}^J \epsilon_J~.\label{dilatino}
\eea
Here  $\Gamma_\mu$, $\mu=1,\ldots,6$, are taken to be Hermitian and generate the Clifford 
algebra $\mathrm{Cliff}(6,0)$ in an orthonormal frame. We have defined the chirality operator
 $\Gamma_7 = \ii \Gamma_{123456}$, which satisfies $(\Gamma_7)^2=1$. 
The covariant derivative acting on the spinor is $D_\mu\epsilon_I={\nabla}_\mu\epsilon_I+\frac{\ii}{2} A^i_\mu(\sigma_i)_I{}^J\epsilon_J$, where ${\nabla}_\mu=\partial_\mu+\frac{1}{4}\Omega_{\mu}^{\ \, \nu\rho}\Gamma_{\nu\rho}$ denotes the Levi-Civita 
spin connection while $\sigma_i$, $i=1,2,3$, are the Pauli matrices.

For simplicity we shall consider Abelian solutions in which 
$A^1_\mu=A^2_\mu=0$, and $A^3_\mu\equiv \mathcal{A}_\mu$, with field strength $\mathcal{F}\equiv \diff \mathcal{A}$. 
Also, as in \cite{Alday:2014bta}, we consider a ``real'' class of solutions
 for which $\epsilon_I$ satisfies the symplectic Majorana condition 
 $\varepsilon_{I}^{\ J}\epsilon_J = \mathcal{C}\epsilon_I^*\equiv \epsilon_I^c$, where 
 $\mathcal{C}$ denotes the charge conjugation matrix, satisfying $\Gamma_\mu^\mt=\mathcal{C}^{-1}\Gamma_\mu \mathcal{C}$. The bosonic fields 
 are all taken to be real, with the exception of the $B$-field which is purely imaginary. 
 With these reality properties one can show that the Killing spinor equation (\ref{KSE}) and dilatino equation 
 (\ref{dilatino}) for $\epsilon_2$  are simply the charge conjugates of the corresponding 
 equations for $\epsilon_1$.
  In this way 
 we  effectively reduce to a single Killing spinor $\epsilon\equiv \epsilon_1$, with $SU(2)_R$ doublet $(\epsilon_1,\epsilon_2)=(\epsilon,\epsilon^c)$. 

\subsection{$SU(2)$ structure}\label{SecSU2}

Consider a Dirac spinor $\epsilon$ in six dimensions, such that $(\epsilon_1,\epsilon_2)=(\epsilon,\epsilon^c)$ 
solves  (\ref{KSE}) and (\ref{dilatino}) above. 
We may construct the following scalar bilinears
\bea
S &\equiv & \epsilon^\dagger\epsilon~, \qquad \tilde{S} \ \equiv \ \epsilon^\dagger\Gamma_7\epsilon~, \qquad 
f \ \equiv \ \epsilon^\mt \epsilon~.
\eea
Here we have chosen a basis for the gamma matrices in which they are purely imaginary and anti-symmetric, 
with charge conjugation matrix $\mathcal{C}=-\ii \Gamma_7$. 
A short computation reveals that
\bea
\diff (Xf) &=& -\ii ( Xf) \mathcal{A}~.
\eea
The integrability condition for this equation immediately implies 
$\mathcal{F}=\diff \mathcal{A}=0$ unless $f\equiv 0$ (notice that 
$X$ is nowhere zero). We will henceforth restrict our analysis to the 
case $f\equiv 0$, which is necessary for a non-trivial R-symmetry gauge field.\footnote{There are nevertheless interesting solutions 
for which $f\neq 0$. In particular the 1/2 BPS solution constructed in \cite{Alday:2014bta} lies in this class.} 

We may then write 
\bea\label{epm}
\epsilon &=& \epsilon_+ + \epsilon_-~,
\eea
where $-\Gamma_7\epsilon_\pm = \pm \epsilon_\pm$, and furthermore the condition $f\equiv 0$ allows us  to introduce \cite{Gauntlett:2004zh}
\bea\label{SU2rotation}
\epsilon_+ &=& \sqrt{S}\cos\alphaangle\, \eta_1~, \qquad \epsilon_- \ = \ \sqrt{S}\sin\alphaangle\, \eta_2^*~.
\eea
Here $\eta_1$, $\eta_2$ are two orthogonal unit norm chiral spinors, so that 
$\eta_1^\dagger\eta_1=\eta_2^\dagger\eta_2=1$ and $\eta_2^\dagger\eta_1=0$. 
These each define a canonical $SU(3)$ structure, and together determine a canonical  $SU(2)$ structure. 
Concretely, in six dimensions such a structure is specified by two one-forms
$K_1$, $K_2$ and a triplet of two-forms $J_i$, $i=1,2,3$, given by
\bea
K_1 - \ii K_2 &\equiv & -\frac{1}{2}\varepsilon^{\alpha\beta}\eta_\alpha^\mt\Gamma_{(1)}\eta_\beta~,\nn\\
J_i &\equiv  & -\frac{\ii}{2}\sigma_i^{\alpha\beta}\eta_\alpha^\dagger \Gamma_{(2)} \eta_\beta~.
\eea
Here we have introduced the notation $\frac{1}{n!}\Gamma_{\mu_1\cdots \mu_n}\diff x^{\mu_1}\wedge \cdots \wedge \diff x^{\mu_n}$, 
where $x^\mu$  are local coordinates. We also define
\bea
\Omega & \equiv & J_2 + \ii J_1~, \qquad J \ \equiv \ J_3~.
\eea
The canonical $SU(2)$ structure is thus determined by $(K_1,K_2,J,\, \Omega)$. 
We note that $K_1$ and $K_2$ are orthonormal one-forms, and both are orthogonal 
to $J$ and $\Omega$, with $J\wedge \, \Omega=0$ and $2J\wedge J = \Omega\wedge\bar{\Omega}$. 

The $SU(2)$ structure $(S,\alphaangle,K_1,K_2,J,\, \Omega)$ that arises naturally from a supersymmetric 
solution is thus related to the canonical $SU(2)$ structure by the square norm $S$ and angle $\alphaangle$, 
via (\ref{SU2rotation}). For completeness we note that $\tilde{S}=-S\cos 2\alphaangle$.

Before proceeding, let us remark that the spinor  $\epsilon$ is charged under the 
Abelian R-symmetry gauge field 
$\mathcal{A}$, and thus it is rotated by a phase under gauge transformations. 
The two-form $\Omega$ is then rotated by the square of this phase. 
As a consequence we more precisely have a $U(2)$ structure, 
as explained in \cite{Alday:2015lta}. Nevertheless, in this paper we will continue 
to refer to this as an $SU(2)$ structure.

\subsection{Differential constraints}

We begin by introducing the one-form bilinear
\bea\label{Kdefinition}
K & \equiv & \epsilon^\dagger \Gamma_{(1)} \epsilon \ = \ S\sin 2\alphaangle\, K_1~.
\eea
Using the Killing spinor equation (\ref{KSE}) and dilatino equation (\ref{dilatino}) 
one can show that $K$ is a Killing one-form, so that the dual vector field 
$\xi\equiv K^{\#}$ is a Killing vector. We may hence introduce a local coordinate $\psi$, 
so that $\xi=\partial_\psi$ and the metric is independent of $\psi$. From (\ref{Kdefinition}) 
it follows that we may write
\bea\label{K1decomp}
K_1 & = & S\sin 2\alphaangle\, (\diff \psi + \sigma)~, 
\eea
where $\mathcal{L}_\xi\sigma = 0 = i_\xi \sigma$. In fact, as shown in appendix \ref{SecDiff}, 
all of the supergravity
fields and $SU(2)$ structure are annihilated by $\mathcal{L}_\xi$, 
with the exception of the complex two-form $\Omega$. The spinor $\epsilon$ 
is a spin$^c$ spinor, charged under the Abelian R-symmetry gauge field 
$\mathcal{A}$, and provided one makes the gauge choice (\ref{Agaugechoice}) 
below then also $\mathcal{L}_\xi\Omega=0$. Thus the vector field $\xi=\partial_\psi$ 
generates a symmetry of the full solution. 

The spinor equations (\ref{KSE}), (\ref{dilatino}) impose further constraints 
on the supergravity fields and $SU(2)$ structure. A more detailed analysis 
may be found in appendix \ref{SecDiff}, while here we simply summarize 
the results. The $B$-field and R-symmetry gauge field strength $\mathcal{F}=\diff\mathcal{A}$ 
may be written as
\bea\label{Bdecomp}
B &=& \ii \, K_1\wedge \left[\frac{3}{\sqrt{2}S\sin 2\alphaangle}\diff (XS) + X^{-2} K_2\right] + B_\perp~,\\\label{Fdecomp}
\mathcal{F} &=& K_1\wedge \frac{\sqrt{2}}{S\sin 2\alphaangle}\diff (XS\cos 2\alphaangle)+ \mathcal{F}_\perp~,
\eea
where $B_\perp$ and $\mathcal{F}_\perp$ have zero interior contraction with $\xi$. 
In particular (\ref{Fdecomp}) allows us to write
\bea\label{Agaugechoice}
\mathcal{A} &=& -\sqrt{2}X\cot 2\alphaangle\, K_1 + \mathcal{A}_\perp~,
\eea
where $i_\xi \mathcal{A}_\perp=0$ and we have made a partial gauge choice for $\mathcal{A}$. We note that
\bea\label{dAperp}
\mathcal{F}_\perp &=& -\sqrt{2}X S \cos 2\alphaangle\, \diff \sigma + \diff \mathcal{A}_\perp~.
\eea
We may similarly write the component of $H=\diff B$ perpendicular to $\xi$ as
\bea\label{Hperptext}
H_\perp &\equiv & \ii \left[\tfrac{3}{\sqrt{2}}\diff (XS) + X^{-2} S\sin 2\alphaangle K_2\right]\wedge \diff \sigma + \diff B_\perp~.
\eea 

Given these definitions, the spinor equations (\ref{KSE}), (\ref{dilatino}) imply the following 
set of differential constraints on the $SU(2)$ structure $(S,\alphaangle,K_1,K_2,J,\, \Omega)$:
\bea
X^2S^2\sin^2 2\alphaangle \, \diff \sigma &=& -\tfrac{2\sqrt{2}}{3}X^{-1}S\cos 2\alphaangle \,  J 
- \ii X^4 S\sin 2\alphaangle\, K_1\hook *H_\perp \nn\\
&&  + \sqrt{2} XS (\cos 2\alphaangle \, \mathcal{F}_\perp + \tfrac{2}{3}\ii B_\perp)~,\nn\\
\diff (X^{-1}S\cos 2\alphaangle\, J) &=& -\tfrac{3}{2\sqrt{2}} \diff[(XS)^2 \diff\sigma] 
+ \ii XS \, \diff B_\perp \nn\\&&
+ \tfrac{\sqrt{2}}{3}\ii X^{-2}S\sin 2\alphaangle \left[K_1\hook *B_\perp 
- K_2\wedge B_\perp\right]~,\nn\\
\diff (X^{-1}SJ) &=& -\sqrt{2}S\sin 2\alphaangle\, J\wedge K_2 - \tfrac{3}{2\sqrt{2}}\cos 2\alphaangle \, \diff [(XS)^2 \diff \sigma]  \nn \\ && + \ii XS \cos 2\alphaangle\, \diff B_\perp
- \tfrac{1}{\sqrt{2}}X^{-2} S \sin 2\alphaangle\, \left[K_1\hook *\mathcal{F}_\perp - K_2\wedge\mathcal{F}_\perp\right]\, ,\nn\\
\diff (S\sin 2\alphaangle\, J\wedge K_2) &=& 0~,\nn\\
D_\perp(X^{-1}S\sin 2\alphaangle\, \Omega) &=& -\sqrt{2}S\Omega\wedge K_2~,\nn\\
S^2 J\wedge \diff\sigma &=& -\sqrt{2}S\cos 2\alphaangle (X+\tfrac{2}{3}X^{-3})\tfrac{1}{2}J\wedge J + 2S K_1\hook *\diff \alphaangle \nn\\
&& +\tfrac{1}{\sqrt{2}}X^{-1}SJ\wedge (\cos 2\alphaangle\, \diff\mathcal{A}_\perp + \tfrac{2}{3}\ii B_\perp)~,\nn\\
S^2 \Omega\wedge \diff\sigma & =&   -2\ii S \diff \alphaangle\wedge K_2 
\wedge \Omega  + \tfrac{1}{\sqrt{2}}X^{-1}S\Omega\wedge (\cos 2\alphaangle\, 
\diff\mathcal{A}_\perp + \tfrac{2}{3}\ii B_\perp)~,\nn\\
0 &=& X^4 K_2 \hook \diff (X^{-3}S\sin 2\alphaangle) + \sqrt{2}S(X^2-\tfrac{2}{3}X^{-2})  \nn\\
&&+\tfrac{1}{\sqrt{2}}S J\hook
(\mathcal{F}_\perp + \tfrac{2}{3}\ii \cos 2\alphaangle\, B_\perp)~.\label{diffconstraints}
\eea
Here the covariant derivative is $D_\perp=\diff + \ii\mathcal{A}_\perp\wedge$, and the interior contraction 
of a $p$-form $\rho$ into a $q$-form $\lambda$ (with $q\geq p$) is the $(q-p)$-form 
$(\rho\hook \lambda)_{\mu_1\cdots \mu_{q-p}}\equiv \frac{1}{p!} \rho^{\nu_1\cdots \nu_p}\lambda_{\nu_1\cdots \nu_p 
\mu_{1}\cdots\mu_{q-p}}$. Notice that the one-form $\sigma$ effectively determines $K_1$ via (\ref{K1decomp}), 
while the supergravity fields enter the equations via $X$, $\mathcal{A}_\perp$ and $B_\perp$. 

\subsection{Sufficiency}\label{SecSuff}

In this section we shall argue  that (\ref{diffconstraints}) are in fact \emph{equivalent} 
to the original spinor equations (\ref{KSE}), (\ref{dilatino}), and moreover as shown in appendix \ref{SecInt} 
these imply 
all but one component of the equations of motion (\ref{FullEOM}), (\ref{Einstein}). 

As in equation (\ref{epm}), we may decompose the Killing spinor as $\epsilon=\epsilon_++\epsilon_-$,  
where $\epsilon_\pm$ have definite chirality under $\Gamma_7$. Each of these defines an $SU(3)$ structure 
in six dimensions, which is equivalent to specifying the 
real two-forms $\mathcal{J}_\pm\equiv -\ii \epsilon_\pm^\dagger \Gamma_{(2)}\epsilon_\pm$ and complex three-forms
 $\Omega_\pm \equiv \epsilon_\pm^\mt\Gamma_{(3)}\epsilon_\pm$. For each choice of $\pm$, there exists 
 a generalized connection with torsion $\nabla_\pm^{(T)}$ which preserves the corresponding structure, {\it i.e.} 
 $\nabla_\pm^{(T)}\epsilon_\pm=0$. One then defines the 
  \emph{intrinsic torsion} as $\tau_\pm\equiv \nabla_\pm^{(T)}-\nabla$, where $\nabla$ is the Levi-Civita connection. 
The exterior derivatives of  $\mathcal{J}_\pm$ and $\Omega_\pm$ determine 
 completely the corresponding intrinsic torsions. One can thus regard the Killing spinor equation
 as an equation that relates the exterior derivatives of $\mathcal{J}_\pm$ and $\Omega_\pm$, 
 on the left hand side of  (\ref{KSE}), to the supergravity fields on the right hand side. 
 Since 
 \bea
 \mathcal{J}_\pm &=& \tfrac{1}{2}S (1\pm \cos 2\alphaangle)(J \mp K_1\wedge K_2)~, \nn\\
 \Omega_\pm &=& \tfrac{1}{2}S (1\pm \cos 2\alphaangle)\, \Omega \wedge (\mp K_1 + \ii K_2)~,
 \eea
 our equations (\ref{diffconstraints}) certainly contain this information, as they imply 
 the exterior derivatives of all $k$-form bilinears, for $k\leq 3$ (this is clear from the 
 analysis in appendix~\ref{SecDiff}). 
 In fact they contain 
 more than this information, as we have also used the dilatino constraint (\ref{dilatino}) 
 to further simplify the equations. 

It thus remains to show that (\ref{diffconstraints}) imply the dilatino equation (\ref{dilatino}). 
First we note that neither $\epsilon_+$ nor $\epsilon_-$ can be identically zero. For 
if $\epsilon_\pm=0$, respectively, then we 
 in fact have an $SU(3)$ structure, rather than $SU(2)$ structure, and 
the bilinear $W\equiv \epsilon^\mt \Gamma_{(3)}\epsilon = \Omega_\mp$
 is the 
corresponding complex three-form.  However, since the left hand side of equation (\ref{biO}) 
of appendix \ref{SecDiff} is identically zero, we would deduce that $\Omega_\mp=0$ and 
hence $\epsilon_\mp=0$. Thus on an open dense subset where $\epsilon_\pm$ 
are both non-zero, we have that $\{\epsilon_\pm, \Gamma_\mu \epsilon_\pm^*\}$ 
span the positive and negative chirality spin bundles $\Spin^\pm$, respectively. 
In order for the dilatino equation to hold, it is therefore sufficient to check that
the contraction of the right hand side of (\ref{dilatino}) 
with $\epsilon_\pm^\dagger$ and $\epsilon_\pm^\mt\Gamma_\mu$ is zero. 
These are equivalent to two scalar and two one-form equations, respectively, that may be 
expressed in terms of bilinears. The corresponding equations may be found in 
appendix \ref{SecDil}. 
It is straightforward, but somewhat tedious, 
to show that these are indeed implied by (\ref{diffconstraints}). 

We thus conclude that (\ref{diffconstraints}) are in fact necessary and sufficient for the 
original spinor equations (\ref{KSE}), (\ref{dilatino}) to hold.

\subsection{Summary}

We have shown that a real supersymmetric solution to Euclidean Romans supergravity, 
with non-trivial Abelian R-symmetry gauge field $\mathcal{A}$, is described by 
an $SU(2)$ structure $(S,\alphaangle,K_1=S\sin 2\alphaangle(\diff \psi+\sigma),K_2,J,\, \Omega)$ with corresponding metric
\bea\label{general6d}
\diff s^2 &=& S^2\sin^2 2\alphaangle (\diff\psi+\sigma)^2 + K_2^2 + g_{SU(2)}~.
\eea
Here we may complete $K_1, K_2$ to an orthonormal frame $\{e^a,e^5\equiv K_1,e^6\equiv K_2\}$, $a=1,\ldots,4$, 
where 
\bea\label{SU2str}
g_{SU(2)} \ = \ \sum_{a=1}^4 (e^a)^2~, \quad J \ = \ e^1\wedge e^2 + e^3\wedge e^4~, \quad 
\Omega \ = \ (e^1+\ii e^2)\wedge (e^3+\ii e^4)~.
\eea
The vector field $\xi=\partial_\psi$ is a Killing vector, and all  supergravity fields
and the $SU (2)$ structure are annihilated by $\mathcal{L}_\xi$ in the gauge for which 
\bea\label{Aagain}
\mathcal{A} &=& -\sqrt{2}X\cot 2\alphaangle\, K_1 + \mathcal{A}_\perp~.
\eea
The Killing spinor equation (\ref{KSE}) and dilatino equation (\ref{dilatino}) 
are then equivalent to imposing the differential constraints (\ref{diffconstraints}) 
on this structure, where $B_\perp$ is the component of the $B$-field with zero 
interior contraction with $\xi$. Moreover, these imply all of the equations 
of motion (\ref{FullEOM}), (\ref{Einstein}) provided we also impose 
\bea
\label{extra}
&&0 \ =\ X^4 S\sin 2\alphaangle \, \diff \sigma \wedge (K_1\hook *\, \ii  H_\perp) + \diff \left[\frac{X^4}{S\sin 2\alphaangle}K_1\hook * \diff (X^{-2} S \sin 2\alphaangle \, K_2)\right]
 \nn\\
 && +\tfrac{2}{9}B_\perp\wedge B_\perp + \tfrac{1}{2}\mathcal{F}_\perp\wedge \mathcal{F}_\perp - 
 \tfrac{4 }{9}X^{-2}K_1\hook  *\left[\frac{3}{\sqrt{2}S \sin 2\alphaangle}\diff (XS) + X^{-2}K_2\right]~.
\eea
This is the component of the $B$-field equation of motion in (\ref{FullEOM}) that has 
zero interior contraction with $\xi$, where recall that $H_\perp$ is defined by (\ref{Hperptext}). 

\section{Applications}\label{SecApplications}

\subsection{Expansion at a conformal boundary}\label{SecExpansion}

In this section we determine the asymptotic form of the $SU(2)$ structure at a conformal boundary. 
The aim is to make contact with the results of \cite{Alday:2015lta}. A similar holographic approach 
to constructing rigid supersymmetric backgrounds in lower dimensions was followed in \cite{Klare:2012gn, Klare:2013dka, Hristov:2013spa}.

Given an
 asymptotically locally AdS solution  we may introduce a radial coordinate $r$ with the conformal boundary located at $r=\infty$. 
The bosonic fields then admit an expansion of the form
\bea
\diff s^2 &=& \frac{9}{2}\frac{\diff r^2}{r^2}+ r^2\left[g^{(0)}_{mn}+\frac{1}{r^2}g^{(2)}_{mn} + \cdots\right]\diff x^m\diff x^n ~,\nn \\
X &=& 1 + \frac{1}{r^2}X_2+\cdots~,\nn\\
B &=& r b - \frac{1}{r^2}\diff r\wedge A^{(0)}+\cdots ~,\nn\\
\mathcal{A} &=& a  +  \cdots~,
\eea
where recall $H =\dd B$ and $\mathcal{F} = \dd \mathcal{A}$. 
The five-dimensional coordinates on the conformal boundary are denoted $x^m$, with $m=1,2,3,4,5$. Some of the terms
{\it a priori} present in these expansions are set to zero by the equations of motion.

In order to determine the corresponding expansion of the $SU(2)$ structure, for this subsection we introduce the following explicit basis 
for Cliff$(6,0)$:
\begin{align}
 \Gamma_{m} \ =\  \left(\begin{array}{cc}0 & \ii \gamma_{m} \\ - \ii \gamma_{m} & 0\end{array}\right) \, , \quad \Gamma_6 \ =\ & \left(\begin{array}{cc}0 & -1_4 \\-1_4 & 0\end{array}\right) \, , \quad 
 \Gamma_7 \ =\  \left(\begin{array}{cc}-1_4 & 0 \\0 & 1_4\end{array}\right) \, , \label{CliffExp}
\end{align}
where $\gamma_{m}$ are a Hermitian basis of Cliff$(5,0)$. Notice that (\ref{CliffExp}) is different to the basis used 
in the rest of the paper (where $\Gamma_\mu$ are purely imaginary), but instead coincides with the 
basis used in \cite{Alday:2015lta}.
The asymptotic form of the metric implies the radial expansion of an orthonormal frame is
\bea
E^6 &=& -\frac{3}{\sqrt{2}}\frac{\diff r}{r}~, \qquad E^m \ = \ r e^m + \cdots \, . \label{FrameExp}
\eea
The Killing spinor then has the following   asymptotic expansion 
\begin{align}
\epsilon \ = \ \sqrt{r} \left(\begin{array}{c} \chi \\  -\ii \chi \end{array}\right) + \frac{1}{\sqrt{r}} \left(\begin{array}{c} \varphi \\ \ii \varphi \end{array}\right) + \cdots \, .\label{KSEExp}
\end{align}
From this, together with $S\equiv \epsilon^\dagger\epsilon$ and the definitions in \eqref{bilinears}, we deduce the following asymptotic expansion for the $SU(2)$ structure:
\begin{eqnarray}\label{expansions}
S &=& 2 {\Sf} (x) \, r + \cdots ~,\nn \\
\alphaangle &=& \frac{\pi}{4} + \frac{\alphaanglef(x)}{r} + \cdots ~, \nn \\
K_1 &=& \Kf_1(x) \, r + \cdots ~, \nn \\
K_2 &=& \Kf_2(x) - \frac{3}{\sqrt{2}}\frac{\dd r}{r}  + \cdots ~, \nn \\
J &=& \Jf(x) \, r^2 + \cdots ~, \nn \\
\Omega &=& \Omegaf(x) \, r^2 + \cdots ~,
\end{eqnarray}
where the ellipses denote subleading terms. Inserting these expansions into (\ref{diffconstraints}) reduces to the following independent 
equations, at leading order in $r$:
\bea
\label{leadingreqs}
\dd \Sf &=& - \tfrac{\sqrt{2}}{3}  \left( \Sf \Kf_2 + \ii {\Sf \Kf_1}\hook b \right) \, , \nn \\
\dd ( \Sf \alphaanglef ) &=& - \tfrac{1}{2\sqrt{2}} {\Sf \Kf_1} \hook \dd a \, , \nn \\
\dd ( \Sf \Kf_1 ) &=& \tfrac{2\sqrt{2}}{3} \left[ 2 \alphaanglef \Sf \Jf + \Sf \Kf_1 \wedge \Kf_2 + \ii \Sf b - \tfrac{\ii}{2}{\Sf \Kf_1}\hook ( * b ) \right] \, , \nn \\
\dd ( \Sf \Kf_2 ) &=& \ii {\Sf \Kf_1}\hook \dd b  -\ii {\Sf \Kf_1}\hook \dd (\log \Sf) b \, , \nn \\
\dd ( \Sf \Jf ) &=& - \sqrt{2} \Kf_2 \wedge ( \Sf \Jf ) \, , \nn \\
\dd ( \Sf \Omegaf ) &=& - \ii \left( a - 2 \sqrt{2} \alphaanglef \Kf_1 - \ii \sqrt{2} \Kf_2 \right) \wedge ( \Sf \Omegaf ) \, . 
\eea
Here $*$ denotes the Hodge duality operator for the boundary metric $g^{(0)}$. We also note that 
the flux equation of motion (\ref{extra}) does not impose an independent constraint at leading order. The set of equations (\ref{leadingreqs}) is precisely the starting point for the purely field theory analysis of rigid supersymmetric five-manifold backgrounds carried out in \cite{Alday:2015lta}.

\subsection{BPS Wilson loops}\label{SecWilson}

The expectation value of Wilson loops in $USp(2N)$ SCFTs have been computed when the gauge theory is placed on the round five-sphere \cite{Assel:2012nf} or $SU(3)\times U(1)$ squashed fives-spheres \cite{Alday:2014bta}. Romans supergravity solutions dual to these backgrounds have also been constructed and successfully compared with the large $N$ gauge theory results. In this section we compute the regularised string action dual to the Wilson loops for \emph{any} Romans solution with ball topology and $U(1)^3$ symmetry, confirming one of the conjectures made by the authors in \cite{Alday:2014bta}.

As shown in \cite{Alday:2014bta}, the relevant string action is
\bea\label{stringaction}
S_{\mathrm{string}} &=& \int_{\Sigma_2}X^{-2}\mathrm{vol}_2 + \ii B - \frac{3}{\sqrt{2}}\mathrm{length}(\partial \Sigma_2)~,
\eea
where the boundary counterterm regularizes the divergence arising from the infinite boundary length. We begin by writing
\bea
B & \equiv & B_1\wedge K_1 + B_\perp~.
\eea
Comparing to \eqref{Bdecomp}  we see that
\bea
X^{-2}K_2 &=& -\frac{3}{\sqrt{2}S\sin 2\alphaangle}\diff (XS)+\ii B_1~.
\eea
It is natural to define the radial coordinate
\bea
\rSUSY &\equiv & XS~.
\eea
Then
\bea\label{subK2}
X^{-2}K_2 &=& -\frac{3X}{\sqrt{2}\sin 2\alphaangle}\frac{\diff \rSUSY}{\rSUSY}+\ii B_1~.
\eea
Notice that in general $B_1$ has a component in the $\diff \rSUSY$ direction, and also 
$\diff \rSUSY$ is not orthogonal to $J$ and $\Omega$. However, we may still consider substituting 
(\ref{subK2}) into the bilinears, at the expense of introducing the unknown $B_1$. 
From the point of view of asymptotically locally AdS solutions this is natural, since 
to leading order at large $\rSUSY$ we see from (\ref{expansions}) that $K_2$ is in the $\diff \rSUSY$ direction. Let us next wedge (\ref{subK2}) with $K_1$. This reads
\bea\label{exact}
X^{-2}K_1\wedge K_2+  \ii (B-B_\perp) &=& \frac{3}{\sqrt{2}}\diff \rSUSY\wedge (\diff \psi+\sigma)~.
\eea
The left hand side is precisely the (unregularized) action of a string wrapping the $K_1$--$K_2$ direction, while the right hand side is exact 
on the string worldsheet. 
In appendix \ref{SecCarolina} we show that such a string is supersymmetric.
Notice that 
\bea
\|\partial_\psi\| &=& S\sin 2\alphaangle  \ = \  \rSUSY X^{-1}\sin 2\alphaangle \ =\ \rSUSY + O(1/\rSUSY)~.
\eea
Here we have used the asymptotic expansions in section \ref{SecExpansion}. 
Since the string wraps the $\partial_\psi$ direction, the boundary length is 
\bea
\mathrm{length}(\partial \Sigma_2) \ = \ \|\partial_\psi\| \int_{S^1}\diff \psi~.
\eea
Integrating by parts the bulk action in (\ref{stringaction}), 
we see that the boundary counterterm  simply cancels 
against the bulk contribution at infinity, leaving
\bea\label{Sstring}
S_{\mathrm{string}} &=& -\frac{3}{\sqrt{2}}\rSUSY_{\mathrm{origin}} \int_{S^1}\diff\psi~,
\eea
where
\bea
\rSUSY_{\mathrm{origin}} &=&  (XS)\mid_{\mathrm{origin}}~.
\eea
Here $\rho\in [\rho_{\mathrm{origin}},\infty)$.
We next claim that for a solution with ball topology 
and $U(1)^3$ isometry
\bea\label{XS0}
(XS)\mid_{\mathrm{origin}} &=& \frac{b_1+b_2+b_3}{\sqrt{2}}~.
\eea
Here we write the supersymmetric Killing vector as
\bea\label{psib}
\partial_\psi &=& \sum_{i=1}^3 b_i \partial_{\varphi_i}~,
\eea
where $\varphi_i$, $i=1,2,3$, have period $2\pi$,
and the orientations (and hence signs) will be fixed shortly.
Combining (\ref{XS0}) with (\ref{Sstring}) for a Wilson loop 
wrapping the $\varphi_i$ circle we obtain
\bea
S_{\mathrm{string}} &=& -9\pi \frac{b_1+b_2+b_3}{3b_i}~,
\eea
where $\int_{S^1}\diff\psi = 2\pi/b_i$.
This is precisely the Wilson loop conjecture made by the authors in \cite{Alday:2014bta}. 

Thus it remains to prove (\ref{XS0}). Geometrically, the $b_i$ arise as the skew 
eigenvalues of the two-form $\diff K$ at the origin (recall that $K = S\sin 2\alphaangle \, K_1$ is a Killing one-form).
 That is, raising an index of $\diff K$ to obtain a 
skew-symmetric $6\times 6$ matrix in an orthonormal frame, at the origin we have
\bea\label{dKmat}
(\diff K)\mid_{\mathrm{origin}} &=& \left(\begin{array}{ccc}R_1 & 0 & 0 \\ 0 & R_2 & 0 \\ 0 & 0 & R_3\end{array}\right)~, \qquad 
R_i \ = \ \left(\begin{array}{cc}0 & -b_i \\ b_i & 0\end{array}\right)~.
\eea
This follows from a simple local calculation. Specifically, 
at the origin we may introduce three sets of polar coordinates $(\rho_i,\varphi_i)$, $i=1,2,3$, 
and write the leading order flat metric as
\bea
\diff s^2_{\mathrm{flat}} &=& \sum_{i=1}^3\diff\rho_i^2 + \rho_i^2\diff\varphi_i^2~.
\eea
One can then compute $\dd K$ at the origin using this local metric, where $K=\sum_{i=1}^3b_i\rho_i^2\diff\varphi_i$ is the 
dual one-form to $\partial_\psi$. In the orthonormal frame 
\bea
e_{2i-1} &=& \diff\rho_i~, \qquad e_{2i} \ = \ \rho_i\diff\varphi_i~, \qquad i\ =\ 1,2,3~,
\eea
at the origin this gives precisely (\ref{dKmat}). Our solution is also equipped 
with a six-dimensional almost complex structure, which as a two-form reads
\bea
\mathcal{J} &=& K_1\wedge K_2 + J~.
\eea
In the same frame this reads
\bea
\mathcal{J} &=& \left(\begin{array}{ccc}\varepsilon & 0 & 0 \\ 0 & \varepsilon & 0 \\ 0 & 0 & \varepsilon\end{array}\right)~, \qquad 
\varepsilon \ = \ \left(\begin{array}{cc}0 & -1 \\1  & 0\end{array}\right)~.
\eea
Thus $\mathcal{J}(e_1)=e_2$, {\it etc}. Notice this fixes the orientations of the $\varphi_i$. Then
\bea\label{JK}
\mathcal{J}\hook \diff K\mid_{\mathrm{origin}} &=& 2(b_1+b_2+b_3)~.
\eea

Let us now look at computing the same quantity using the bilinear equations. 
We have
\bea\label{K1dK}
K_1\hook \diff K &=& \frac{1}{X^2}\left[-\frac{1}{S\sin 2\alphaangle}\diff (X^2S^2\sin^22\alphaangle) 
+ 2XS\sin 2\alphaangle \, \diff X\right]~.
\eea
$K$ has norm $S\sin 2\alphaangle$, which by definition is zero at the origin. 
Contracting  $K_2$ into (\ref{K1dK}) and 
restricting to the origin we hence find
\bea\label{d2alphaorigin}
(K_1\wedge K_2)\hook \diff K\mid_{\mathrm{origin}} &=& -2K_2\hook \diff(S\sin 2\alphaangle)\mid_{\mathrm{origin}}~,
\eea
where we have assumed that $X$ is regular at the origin (and we shall make similar regularity assumptions 
for other fields in what follows).
We next compute
\bea
J\hook \diff K &=& (S\sin 2\alphaangle)^2 J\, \hook \diff\sigma~,
\eea
which thus tends to zero at the origin. 
Finally contracting $K_2$ into (\ref{Omegabil}), and restricting to the origin, 
we find
\bea
K_2\hook \diff (S\sin 2\alphaangle)\mid_{\mathrm{origin}} &=& -\sqrt{2}(XS)\mid_{\mathrm{origin}}~.
\eea
Combined with (\ref{d2alphaorigin}), this shows that
\bea
\mathcal{J}\hook \diff K\mid_{\mathrm{origin}} &=& 2\sqrt{2}(XS)\mid_{\mathrm{origin}}~, 
\eea
which together with (\ref{JK}) proves (\ref{XS0}).

\subsection{Squashed Sasaki-Einstein solutions}\label{SecSquashed}

The system of equations for the $SU(2)$ structure in section \ref{SecConditions} is  too complicated to solve 
in general; to find solutions one needs to make some additional assumptions. In this section 
we consider an ansatz that naturally generalizes the 1/4 BPS  solutions (and their 1/2 BPS limit) found in 
\cite{Alday:2014bta}.

We begin by making the following ansatz for the supergravity fields\footnote{Recall that the formula (\ref{Aagain}) 
for
the gauge field    $\mathcal{A}$ requires a specific gauge choice. However,  in \cite{Alday:2014bta} this was presented 
in a different gauge. This accounts for the factor of $-3\diff\psi$ in (\ref{KEansatz}). }
\bea\label{KEansatz}
\diff s^2 &=& \alpha^2(r)\diff r^2 + \gamma^2(r)(\diff \psi +\sigma)^2 + \beta^2(r)\diff s^2_{\mathrm{KE}}~,\nn\\
 B &=& p(r)\diff r\wedge (\diff \psi+\sigma) + \tfrac{1}{2}q(r)\diff \sigma~,\nn\\
 \mathcal{A} &=& f(r)(\diff \psi +\sigma)-3\diff \psi~,\nn\\
 X &=& X(r)~.
\eea
Here we take $\diff s^2_{\mathrm{KE}}$ to be a four-dimensional positively curved K\"ahler-Einstein metric, 
so that a constant $r$ hypersurface is a squashed Sasaki-Einstein five-manifold. Concretely,
this means that $\diff\psi+\sigma$ is a global contact one-form on such a hypersurface, with
\bea
\diff\sigma &=& 2\omega_{\mathrm{KE}}~.
\eea
The ansatz (\ref{KEansatz}) reduces to that in \cite{Alday:2014bta} on taking the K\"ahler-Einstein metric 
to be the Fubini-Study metric on $\mathbb{CP}^2$. Notice also that in writing (\ref{KEansatz}) 
we have taken
the supersymmetric Killing 
vector $\partial_\psi$ to coincide with the Reeb vector field of the squashed Sasaki-Einstein manifold. 

Comparing to section \ref{SecConditions}, and identifying the four-dimensional $SU(2)$ structure metric  in
(\ref{general6d}) with $\beta^2(r)\diff s^2_{\mathrm{KE}}$, allows us to identify
\bea
S\sin 2\alphaangle \ = \ \gamma(r) ~, \quad K_2 \ = \ -\alpha(r)\diff r~, \quad 
\Omega \ = \ \beta^2(r)\Omega_{\mathrm{KE}}~, \quad J \ = \ -\beta^2(r)\omega_{\mathrm{KE}}~,
\eea
where $\Omega_{\mathrm{KE}}$ satisfies\footnote{We have chosen sign conventions so as to agree with those of \cite{Alday:2014bta}.} 
\bea
\diff \Omega_{\mathrm{KE}} &=& -3\ii \sigma\wedge \Omega_{\mathrm{KE}}~.
\eea
We take $S=S(r)$, $\alphaangle=\alphaangle(r)$. From the remaining supergravity fields, we similarly read off
\bea
f(r) &=& 3-\sqrt{2}XS \cos 2\alphaangle~, \qquad \mathcal{F}_\perp \  = \ 2f(r)\omega_{\mathrm{KE}}~, \qquad 
B_\perp \ =\  q(r)\omega_{\mathrm{KE}}~.
\eea
Substituting these into the differential constraints (\ref{diffconstraints}) and flux equation of motion (\ref{extra}) then reduces 
to the following independent ODEs:
\begin{align}\label{KEODEs}
0\ =& \ \ii X^3\left(2 p-q'\right) \sin 2 \alphaangle +\tfrac{2 \sqrt{2}}{3} \alpha \left[\ii q+ (9+{\beta^2}{X^{-2}})\cos 2 \alphaangle\right]- \alpha  XS (3+\cos 4 \alphaangle)~,\nn\\
0\ =&\ \frac{\dd}{\dd r}(X^{-1} S\beta^2 \cos 2\alphaangle)-3\sqrt{2} XS \frac{\dd}{\dd r}(XS)+\ii  XS q'~,\nn\\
0\ =& \ \frac{\dd}{\dd r}(X^{-1} S \beta^2\sin 2\alphaangle)-\sqrt{2}S \alpha \beta^2~,\nn\\
0\ =&\ -2XS+3\sqrt{2} \cos 2\alphaangle + \ii \tfrac{\sqrt{2}}{3} q + \tfrac{1}{\sqrt{2}}\left(\tfrac{2}{3} X^{-2}+X^2 \right) \beta^2\cos 2\alphaangle - \beta^2X \alpha^{-1}\alphaangle'~,\nn\\
0\ =&\ -\sqrt{2} \alpha S  \left[ \left(3 X^4+1\right)\beta^2+18 X^2\right]\sin 2 \alphaangle+ X^2 \left(12  XS \beta '+\sqrt{2} \ii p \beta\right)\beta~,\nn\\
0\ =&\ -\frac{p \beta^4 \csc 2 \alphaangle}{\alpha  X^2S}+6 \sqrt{2} q  XS-\ii q^2-\frac{6\sqrt{2}\ii  S \beta^2 \cos 2 \alphaangle}{X}+18 \ii  X^2S^2-81 \ii~.
\end{align}
Notice that as a consequence of 
parametrization invariance one is free to specify the function $\beta=\beta(r)$. 
Hence (\ref{KEODEs}) are six coupled ODEs for the six functions $(X,S,\alphaangle,\alpha,p,q)$. 
Furthermore, notice that they are independent of the choice of K\"ahler-Einstein metric, 
and are thus equivalent to the equations studied in  \cite{Alday:2014bta}. 
In the latter reference we constructed a two-parameter family of 1/4 BPS 
solutions, as a series expansion both around the conformal boundary 
at $r=\infty$, and as an expansion around Euclidean AdS. 
Specifically, the parameters are 
\bea\label{twoparameters}
f_0 \ \equiv \ \left.f(r)\right|_{\mathrm{boundary}}~, \qquad s^{-1} \ \equiv \ \left.\frac{\gamma(r)}{\beta(r)}\right|_{\mathrm{boundary}}~.
\eea
We hence automatically construct new solutions, with  an arbitrary squashed Sasaki-Einstein 
five-manifold, with squashing parameter $s$, as conformal boundary. 
Setting $s=1$ and $f_0=0$, the conformal boundary is a Sasaki-Einstein manifold with metric $\diff s^2_{\mathrm{SE}}=(\diff\psi+\sigma)^2+\diff s^2_{\mathrm{KE}}$, 
and in the bulk the only non-trivial field is the metric, which is a ``hyperbolic cone'' 
\bea\label{hypcone}
\diff s^2_6 &=& \frac{\diff r^2}{1 + \frac{2}{9}r^2}+  r^2 \diff s^2_{\mathrm{SE}}~.
\eea
When $\diff s^2_{\mathrm{SE}}$ is the round five-sphere this is simply Euclidean AdS$_6$, while 
more generally (\ref{hypcone}) has an isolated Calabi-Yau cone singularity at $r=0$. 
The solutions with general $s$ and $f_0$ have the same behaviour near the 
tip of the cone/origin, and thus in general these supergravity solutions have 
a Calabi-Yau singularity. Nevertheless, this singularity 
does not lead to any  UV divergences in the holographic free energy or 
Wilson loop VEVs. 
Although we were unable to solve the system (\ref{KEODEs}) analytically, 
see the end of section~\ref{SecAnal} for further discussion.

Any solution to Romans $F(4)$ supergravity uplifts to a solution of massive
type IIA supergravity, as a warped product $M_6\times S^4$ \cite{Cvetic:1999un}. For an 
asymptotically locally AdS solution $M_6$, 
these 
are expected to be the gravity duals to a certain family of $USp(2N)$ 
gauge theories, defined on the conformal boundary  of $M_6$. The gauge 
theories arise from a system of
$N$
D4-branes, $N_f$ of D8-branes and an orientifold plane. This 
data is captured in the six-dimensional effective Newton
constant \cite{Jafferis:2012iv}
\bea
G_N &=& \frac{15\pi \sqrt{8-N_f}}{4\sqrt{2}N^{5/2}}~.
\eea
Recall that the two-parameter family of solutions constructed in this section 
reduce to the 1/4 BPS family in \cite{Alday:2014bta} when the K\"ahler-Einstein 
metric is taken to be the Fubini-Study metric on $\mathbb{CP}^2$. 
The computation of the holographic free energy then very closely 
follows that in  \cite{Alday:2014bta}. The upshot is that
\bea\label{free14}
\mathcal{F}_{\mathrm{gravity}} \ = \ I_{\mathrm{renormalized}} \ = \ 
-\frac{27}{4\pi G_N}\cdot \vol(\mathrm{SE})~,
\eea
is independent of the two parameters $s$ and $f_0$. 
Notice that the volume $\vol(\mathrm{SE})$ appearing in (\ref{free14}) 
is that of the Sasaki-Einstein metric, which is the conformal boundary 
metric when $s=1$, even though (\ref{free14}) holds for all $s$. 

\subsubsection*{Comparison to field theory}

We would like to compare (\ref{free14}) with the corresponding 
large $N$ field theory calculation. This involves computing 
the localized partition function of the $USp(2N)$ gauge theories 
on a squashed Sasaki-Einstein background, and taking the
$N\rightarrow \infty $ limit. In \cite{Qiu:2014oqa} 
the perturbative partition function of an arbitrary 
$\mathcal{N}=1$ supersymmetric 
gauge theory was computed on a general $U(1)^3$--invariant 
Sasaki-Einstein five-manifold. For a gauge theory  with
 gauge group $G$ and 
a matter hypermultiplet in an arbitrary representation $\mathcal{R}$, the 
localized perturbative partition function is 
\be \label{pertparttoricSE}
{Z}_{\mathrm{pert}}^{\mathrm{SE}} \ = \ \int_\mathfrak{t} \diff a \, \ex^{- S_\mathrm{cl}} \, \frac{\prod_\alpha S^{\mathrm{SE}}_3 [\, \ii \alpha(a)\,  ;\, {\vec\reeb}\, ]}{\prod_\rho S^{\mathrm{SE}}_3[\, \ii \rho(a) +\tfrac{3}{2}\,  ;\,  {\vec\reeb}\, ]} \,.
\ee
The integration in $a$ is over the Cartan $\mathfrak{t}$ of the gauge group. 
The products are over roots $\alpha$ of $G$ and weights $\rho$ of the representation $\mathcal{R}$, and we have denoted by $S_\mathrm{cl}$ the classical action evaluated on the localization locus. 
Furthermore $S_3^{\mathrm{SE}}[\, x\, ;\, \vec{\reeb}\, ]$ is a generalized version of the triple-sine function
\be\label{SSE}
 S^{\mathrm{SE}}_{3}[\, x\, ;\, \vec{\reeb}\, ] \ \equiv \ \prod_{\vec{m}} \, (\vec{m}\cdot \vec{\reeb}+x) (\vec{m}\cdot \vec{\reeb}+\vec{\xi} \cdot \vec{\reeb} - x) \,.
\ee
Here $\vec{m}=(m_1,m_2,m_3)$ runs over the charge lattice of holomorphic functions on the Calabi-Yau cone over the Sasaki-Einstein 
five-manifold, where $m_i$ is the charge under the $i$th $U(1)$ symmetry. Furthermore, we have written the  supersymmetric 
(Reeb) vector field as
\bea
\xi &=& \sum_{i=1}^3 \xi_i \partial_{\varphi_i}~,
\eea
where $\vec{\reeb}=(\xi_1,\xi_2,\xi_3)$ and $\partial_{\varphi_i}$ generate the $U(1)^3$ isometry.
For example, for the round $S^5$ the Calabi-Yau cone is simply $\C^3$, with a basis of holomorphic 
functions $z_1^{m_1}z_2^{m_2}z_3^{m_3}$, where $m_i\in \Z_{\geq 0}$. In this case, 
(\ref{SSE}) reduces to the standard triple-sine function.

We are interested in evaluating  (\ref{pertparttoricSE}) for the $USp(2N)$ gauge theories,
in the large $N$ limit. This involves 
the asymptotics of the hypermultiplet and vectormultiplet contributions computed in~\cite{Qiu:2014oqa}:
\begin{align}
\log S^{\mathrm{SE}}_3 [\, x\, ; \, \vec{\reeb}\, ]\  & \sim \  - \ii \pi \, \mbox{sgn}(\mbox{Im}\, x) \left[ \left( \frac{x^{3}}{6} +\frac{3 x}{4} \right)  \frac{\textrm{vol}(\mathrm{SE})}{\pi^3} + \frac{x}{24\pi} \sum_{I}\beta_I  \right] \,, \nn\\
\log S^{\mathrm{SE}}_3[\, x+\tfrac{3}{2}\, ;\, \vec{\reeb}\, ] \  & \ \sim \  \ \, \ii \pi \, \mbox{sgn}(\mbox{Im}\,  x) \left[ \left( \frac{x^{3}}{6 } -\frac{ 3x}{8} \right) \frac{\textrm{vol}(\mathrm{SE})}{\pi^3}+ \frac{x}{24\pi}\sum_{I}\beta_I \right]\,.\label{hyperasy}
\end{align}
Here $\beta_I$ are certain parameters defined in \cite{Qiu:2014oqa}, which will not enter the final result.\footnote{$\beta_I$ is 
the length of the $I$th closed Reeb orbit.} We may then compute the leading contribution to the partition function at large $N$  using a saddle point method.
 One specifies 
 an element of the Cartan subalgebra of $USp(2N)$ by its eigenvalues $\{\lambda_1 , \ldots , \lambda_N \}$. 
In the large $N$ saddle point  
 these behave as $\lambda_n \sim N^{1/2} x_n $. One then introduces an eigenvalue density 
\bea
\rho(x) \ = \ \frac{1}{N} \sum_n \delta (x-x_n) \, ,\eea
which has support on a finite interval $[0,x_\star]$. Solving the saddle point approximation to the above matrix model, we find
\be
 \rho(x) \ = \ \frac{4 (8-N_f) x}{9} \,, \quad \text{and} \quad x_{\star} \ = \ \frac{3}{\sqrt{2} \sqrt{8-N_f}}\,,
\ee
which leads to the final result for the large $N$ free energy
\be\label{Fgauge}
 \mathcal{F}_{\mathrm{gauge \, theory}} \ = \ - \frac{9\sqrt{2}}{5\pi^2 \sqrt{8-N_f} } \vol(\mathrm{SE}) N^{5/2} + o( N^{5/2}) \,.
\ee
This precisely agrees with  (\ref{free14}).

The field theory computation above is for the Sasaki-Einstein conformal boundary, with $s=1$ and $f_0=0$. 
On the other hand, in \cite{Alday:2015lta} we conjectured that the partition function 
should depend only on the holomorphic foliation generated by the Killing vector $\xi$. 
Since this is independent of $s$ and $f_0$, this conjecture implies that (\ref{Fgauge})
holds for the entire two-parameter family of 1/4 BPS backgrounds. Since
(\ref{Fgauge})  agrees with (\ref{free14}), this lends credence to the conjecture.
We also regard this as evidence that the 1/4 BPS family of supergravity backgrounds 
is the correct holographic dual, in spite of the Calabi-Yau singularity at the origin.

\subsubsection*{BPS Wilson loops}

Finally, let us discuss the computation of the VEV of BPS Wilson loops on both sides 
of the correspondence. Following a similar computation to that in \cite{Alday:2014bta}, 
in the large $N$ matrix model for the gauge theory this is given by
\bea
\langle \, W \, \rangle &=& \int_0^{x_\star} \ex^{\lambda(x)\beta_I}\, \rho(x)\diff x~,
\eea
where $\beta_I$ is the length of the closed Reeb orbit wrapped by the Wilson loop.\footnote{Recall that the computation 
of \cite{Qiu:2014oqa} is valid for a $U(1)^3$-invariant Sasaki-Einstein manifold, 
for which the index $I$ runs over the rays of the corresponding polyhedral cone.}
At large $N$ one hence obtains
\bea
\log \, \langle \, W \, \rangle &=& x_\star\beta_I N^{1/2} +o(N^{1/2})~.
\eea
On the other hand, in the dual supergravity solution this corresponds 
to a fundamental string wrapping the circle of length $\beta_I$, together 
with the radial direction $r$. We find that the regularized action is
\bea
S_{\mathrm{string}} &=& -\frac{3}{\sqrt{2}\sqrt{8-N_f}}\beta_I N^{1/2}~.
\eea
This should be identified with $-\log \, \langle \, W \, \rangle$ in field theory, 
and we find perfect agreement.

\subsection{Analytic 3/4 BPS solution}\label{SecAnal}

In this section we give some details of a new analytic supersymmetric solution to Euclidean six-dimensional Romans supergravity. This corresponds to the $3/4$ BPS squashed sphere, constructed as a perturbation expansion in \cite{Alday:2014bta}. As shown in  \cite{Alday:2014bta} an interesting family of solutions arises by considering the following $SU(3) \times U(1)$ symmetric ansatz for the supergravity fields
\begin{eqnarray}\label{ansatzapp}
\dd s^2_6 &=& \alpha^2(r)\dd r^2+\gamma^2(r)(\dd\tau+C)^2+\beta^2(r)\Big[\dd \sigma^2 + \frac{1}{4}\sin^2\sigma(\dd \theta^2+\sin^2\theta \dd \varphi^2)\nn\\
&&+\frac{1}{4}\cos^2\sigma\sin^2\sigma (\dd \beta+\cos\theta \dd\varphi)^2\Big]~,\nn\\
B&=& p(r)\dd r\wedge (\dd\tau+C) + \frac{1}{2}q(r)\dd C~,\nn\\
A^i &=& f^i(r)(\dd\tau +C)~,
\end{eqnarray}
where
\bea
C & \equiv & -\frac{1}{2}\sin^2\sigma (\diff \beta + \cos\theta \diff\varphi)~,
\eea
together with $X=X(r)$. The equations of motion for the background $SU(2)_R$ gauge field imply
\begin{equation}
f^i(r) \ = \ \kappa_i f(r)~.
\end{equation}
The equations for the other fields then depend only on the $SU(2)\sim SO(3)$ invariant $\kappa_1^2+\kappa_2^2+\kappa_3^2$, which we can set to one by rescaling $f(r)$. The set of equations for the fields involved in the ansatz  have been listed in the appendix B to  \cite{Alday:2014bta}. In addition, if the solution is supersymmetric there exists a Killing spinor. For the case of the $3/4$ BPS solution the Killing spinor depends on four extra functions, denoted $k_i(r)$, $i=1,2,3,4$ in \cite{Alday:2014bta}, which, together with the fields above, satisfy first order constraints as a result of supersymmetry. Although, as shown in this paper, these constraints are equivalent to the original equations of motions (upon supplementing them with one extra second order equation), we found them more convenient in order to find an analytic form for the solution. 

The solution depends on a single parameter $s$, the squashing parameter, but it is convenient to parametrize it in terms of $b_1=1+\sqrt{1-s^2}$ and $b_2=1-\sqrt{1-s^2}$, introduced in \cite{Alday:2014bta}.  The high amount of supersymmetry implies a large number of constraints (many of them algebraic) which can be used to eliminate all the fields in favour of $k_2(r),k_3(r),X(r)$ and $\beta(r)$. For instance 
\begin{eqnarray}
k_1(r)&=& b_2(b_1+b_2)\frac{k_2(r) \beta(r)}{b_2 k_2^2(r)+b_1 k_3^2(r)} ~,\nn \\
k_4(r) &=& b_1(b_1+b_2)\frac{k_3(r) \beta(r)}{b_2 k_2^2(r)+b_1 k_3^2(r)} ~,\nn\\
\gamma(r)&=& (b_1+b_2) \frac{k_2(r) k_3(r) \beta(r)}{b_2 k_2^2(r)+b_1 k_3^2(r)}~,
\end{eqnarray}
while the expressions for the remaining fields are more complicated.  As a consequence of reparametrization invariance we can demand that $k_2,k_3$ and $X$ depend on $r$ only through $\beta(r)$. It is then convenient to introduce a new variable $\zeta$:
\begin{equation}
(b_1+b_2)\sqrt{b_1 b_2}\, \beta(r) \ \equiv \ \zeta~.
\end{equation}
The remaining equations can be used to eliminate further fields and we end up with a single equation for $v(\zeta) \equiv \zeta^2 X^2(\zeta)$: 
\begin{equation}
\label{final3/4}
v'(\zeta) \ = \ 4\zeta^3\frac{(b_1+b_2)^2(b_1+2 b_2)^2+2 v(\zeta)}{\zeta^4+3 (b_1+b_2)^3(b_1+2b_2)v(\zeta)+3 v^2(\zeta)}~,
\end{equation}
which can be simply solved, for instance, with Mathematica. Equation (\ref{final3/4}) has two inequivalent solutions, each of them depending on a constant of integration. Of those only one has the correct boundary condition at infinity $v(\zeta) = \zeta^2+\cdots$. The constant of integration can then be fixed by requiring regularity at the origin for $X(\zeta)$, which implies $v(0)=0$. This fixes the solution uniquely. Although the explicit solution is too cumbersome to be written here, we give the expansion of $X(\zeta)$ for small and large values of $\zeta$:
\begin{eqnarray}
X(\zeta) &=& \left(\frac{2(b_1+2b_2)}{3(b_1+b_2)} \right)^{1/4}+\cdots,\qquad \qquad \qquad \qquad ~~~~~~\zeta \ll 1~,\\
X(\zeta)&=& 1 -\frac{(b_1-b_2)(b_1+b_2)^2(b_1+2b_2)}{4} \frac{1}{\zeta^2} \nn \\
&&+\frac{(b_1-b_2)(b_1+b_2)^3(b_1+2b_2)^2}{2\sqrt{2}} \frac{1}{\zeta^3} +\cdots,\quad ~~~~~~\zeta \gg 1~.
\end{eqnarray}
For instance, these expansions allow us to fix the parameter $\kappa$ introduced in \cite{Alday:2014bta}. We obtain
\bea
\kappa \ = \ \frac{2 \sqrt{2}(3-\sqrt{1-s^2})^2(1-s^2+\sqrt{1-s^2})}{27 \sqrt{3} s^5}~.
\eea
Finally, let us remark that although cumbersome, the solution contains only roots and rational functions. 

\subsubsection*{Comments on the 1/4 and 1/2 BPS solutions}

The 1/4 BPS squashed sphere solution considered in \cite{Alday:2014bta} is much harder to obtain, the reason being the smaller degree of supersymmetry. More precisely, the Killing spinor now depends on only two new functions $k_1(r)$ and $k_2(r)$, but the number of constraints is much smaller. A related issue is that now there are no natural ``constants of motion" such as $b_1$ and $b_2$ to parametrize the solution with. Proceeding as before one can write two (third order and very cumbersome!) equations for two of the fields, for instance $X(\zeta)$ and $f(\zeta)$. After requiring regularity at the origin this should lead to a two-parameter family of solutions ($s$ and $f_0$ introduced in (\ref{twoparameters})). These equations, however, are very complicated and we haven't managed to solve them exactly. Before proceeding, two comments are in order: first, these two equations can be solved in different limits, and reproduce the $1/4$ BPS solution in the limits studied in  \cite{Alday:2014bta}. Furthermore, in order to obtain these two equations it is necessary to supplement the bilinear equations with (\ref{extra}). Otherwise, we would obtain only one equation for two fields. This example shows that the differential constraints (\ref{diffconstraints}) do indeed need to be supplemented by
 equation (\ref{extra}). 

We can also consider the special case $f(\zeta)=0$. In this case the 1/4 BPS solution reduces to the 1/2 BPS solution studied in \cite{Alday:2014bta}. Although not covered by our analysis in this paper because the bilinear $\epsilon^\mathrm{T} \epsilon \neq 0$, the 1/2 BPS solution is a limit of the 1/4 BPS solution, where one of the two parameters, namely $f_0$, vanishes. The final equation for $X(\zeta)$, with $\beta(r )\equiv \zeta$ is still rather involved, but it can be solved analytically in an interesting limit. Denoting $X(0)= x_0$ one can explicitly check the solution takes the following form
\begin{equation}
v(\zeta) \ = \  v_0(x_0 \zeta)+ \frac{1}{x_0^4} v_1(x_0 \zeta)+\cdots~,
\end{equation}
where recall $v(\zeta) \equiv \zeta^2 X^2(\zeta)$ and $v_0(y)$ satisfies a simple equation
\begin{equation}
v_0''(y) \ =\   3\frac{v_0'(y)}{y}-\frac{(6+v_0(y))v_0'(y)^2}{6v_0(y)}~,
\end{equation}
whose solution with correct boundary conditions is
\begin{equation}
v_0(y)\ =\   1+{\cal W}\left(\frac{y^4-72\, \ex}{72\, \ex} \right)~.
\end{equation}
Here ${\cal W}(z)$ is the Lambert W function or product logarithm, namely ${\cal W}(z) \ex^{{\cal W}(z)}=z$. Hence, as opposed to the 3/4 BPS solution, this solution contains special functions. 


\section{Discussion}\label{SecDiscussion}

In this paper we have presented a
systematic study of supersymmetric solutions to 
six-dimensional Euclidean Romans supergravity. These are characterized 
by an $SU(2)$ structure. We then used these results to study a number 
of different applications. 

Our results raise a number of interesting questions and directions 
for future work. Firstly, the 
gravity duals to (squashed) Sasaki-Einstein backgrounds we constructed 
have 
isolated Calabi-Yau singularities. However, as we have seen,
the singularity 
does not contribute additional (UV) divergences to 
the free energy and Wilson loop, and moreover the supergravity computations agree
with the gauge theory results. It is thus natural to conjecture 
that these are the correct gravity duals. More precisely, 
although one expects some stringy degrees of freedom 
to be supported at the singularity, we expect that 
these should not contribute to leading order at large $N$.
Notice in any case
that the uplift to massive IIA is also singular (along the internal
$S^4$), even for Euclidean AdS$_6$ 
\cite{Ferrara:1998gv, Brandhuber:1999np}.

Using the technology developed in the paper, 
we have computed the VEV of  the holographic dual of a supersymmetric Wilson loop 
for a general class of solutions, thus proving one of the 
conjectures of \cite{Alday:2014bta}. Another conjecture 
made in that paper makes a specific prediction for the holographic free energy
for the same class of backgrounds. It would be interesting to prove this conjecture.
Note that this computation is more involved than that for the Wilson loop; 
in particular the structure of the counterterms is much more complicated.

Finally, it would be interesting to construct further analytic solutions, including solutions with 
different topology. 


\subsection*{Acknowledgments}

\noindent 
The work of L.~F.~A., M.~F. and P.~R. is supported by ERC STG grant 306260. L.~F.~A. is a Wolfson Royal Society Research Merit Award holder. C.~M.~G. is supported by a CNPq scholarship. J.~F.~S. is supported by the Royal Society. 


\appendix

\section{Useful identities}\label{SecUseful}

From the dilatino equation (\ref{dilatino}) one can derive the following useful identities
\bea
(\partial^\mu X) \epsilon^\dagger [ \mathbb{A} , \Gamma_\mu ]_\mp \epsilon & =&  - \tfrac{\ii}{2\sqrt{2}} \left( X^2 - X^{-2} \right) \epsilon^\dagger [ \mathbb{A} , \Gamma_7 ]_\pm \epsilon + \tfrac{1}{24} X^3 H^{\mu\nu\rho} \epsilon^\dagger [ \mathbb{A} , \Gamma_{\mu\nu\rho} \Gamma_7 ]_\pm \epsilon \nonumber \\
& & + \tfrac{\ii}{12\sqrt{2}}  B^{\mu\nu} \epsilon^\dagger [ \mathbb{A} , \Gamma_{\mu\nu} ]_\pm \epsilon - \tfrac{1}{8\sqrt{2}}  \mathcal{F}^{\mu\nu} \epsilon^\dagger [ \mathbb{A} , \Gamma_{\mu\nu} \Gamma_7 ]_\pm \epsilon \, , \label{DiracDilatino} \\[10pt]
(\partial^\mu X) \epsilon^\mt [ \mathbb{A} , \Gamma_\mu ]_\mp \epsilon & =& - \tfrac{\ii}{2\sqrt{2}} \left( X^2 - X^{-2} \right) \epsilon^\mt [ \mathbb{A} , \Gamma_7 ]_\mp \epsilon + \tfrac{1}{24} X^3 H^{\mu\nu\rho} \epsilon^\mt [ \mathbb{A} , \Gamma_{\mu\nu\rho} \Gamma_7 ]_\pm \epsilon \nonumber \\
& & + \tfrac{\ii}{12\sqrt{2}}  B^{\mu\nu} \epsilon^\mt [ \mathbb{A} , \Gamma_{\mu\nu} ]_\mp \epsilon - \tfrac{1}{8\sqrt{2}}  \mathcal{F}^{\mu\nu} \epsilon^\mt [ \mathbb{A} , \Gamma_{\mu\nu} \Gamma_7 ]_\pm \epsilon \, . \label{ChargeDilatino}
\eea
Here $\mathbb{A}\in \mathrm{Cliff}(6,0)$ is an arbitrary element of the Clifford algebra, 
while $[\, \cdot\, ,\, \cdot\, ]_-$ denotes a commutator and $[\, \cdot\, ,\, \cdot\, ]_+$ denotes an anti-commutator.

\section{Differential conditions for bilinears}\label{SecDiff}

We may introduce the following  bilinears in the spinor $\epsilon$:
{\allowdisplaybreaks
\bea\label{bilinears}
K \ \equiv \  \epsilon^\dagger \Gamma_{(1)}\epsilon & = & S\sin 2\alphaangle\, K_1~,\nn\\
\tilde{K} \  \equiv \ \ii\epsilon^\dagger\Gamma_{(1)}\Gamma_7\epsilon & = & - S \sin 2\alphaangle\, K_2~,\nn\\
Y \ \equiv \  \ii\epsilon^\dagger\Gamma_{(2)}\epsilon & = & S (  \cos 2\alphaangle\,  K_1\wedge K_2 - J)~,\nn\\
\tilde{Y} \  \equiv \ \ii\epsilon^\dagger\Gamma_{(2)}\Gamma_7\epsilon & = & S ( - K_1\wedge K_2 +\cos 2\alphaangle\, J)~,\nn\\
Z \  \equiv \ \epsilon^\mt\Gamma_{(2)}\Gamma_7\epsilon & = & -S\sin 2\alphaangle\, \Omega~,\nn\\
V \ \equiv \  \ii\epsilon^\dagger\Gamma_{(3)}\epsilon & = & -S\sin 2\alphaangle\, K_1\wedge J~,\nn\\
\tilde{V} \  \equiv \ \epsilon^\dagger\Gamma_{(3)}\Gamma_7\epsilon & = & -S\sin 2\alphaangle\, K_2\wedge J ~,\nn\\
W \ \equiv \  \epsilon^\mt\Gamma_{(3)}\epsilon & = & S(-\cos 2\alphaangle\, K_1 + \ii\, K_2)\wedge \Omega~,\nn\\
\tilde{W} \  \equiv \ \epsilon^\mt\Gamma_{(3)}\Gamma_7\epsilon & = & S (K_1 - \ii \, \cos 2\alphaangle\, K_2)\wedge \Omega~.
\eea
}
Here $(K_1, K_2, J, \, \Omega\, )$ is the canonical $SU(2)$ structure defined in section \ref{SecSU2}. 

A straightforward  but lengthy calculation shows that the Killing spinor equation (\ref{KSE}) 
and dilatino equation (\ref{dilatino}) imply the following differential 
constraints on the bilinears in (\ref{bilinears}):
{\allowdisplaybreaks
\bea 
\dd ( X S ) & =&  \tfrac{\sqrt{2}}{3} ( X^{-2} \tilde{K} - \ii K\hook B ) \, , \label{biS} \\
\dd ( X \tilde{S} ) & =&  - \tfrac{1}{\sqrt{2}} K\hook \mathcal{F} \, , \label{bitS} \\
\dd ( X^2 K ) & =&  - \tfrac{2\sqrt{2}}{3} X^{-1} \tilde{Y} - \ii  X^4 K\hook *H - 
\sqrt{2} X (\tilde{S}\mathcal{F} - \ii \tfrac{2}{3}S B)\, , \label{biK} \\
\dd ( X^{-2} \tilde{K} ) & =&  - \ii K\hook H \, , \label{bitK} \\
\dd ( X^{-1} Y ) & =&  - \sqrt{2} \tilde{V} + \ii ( X \tilde{S} ) H + \tfrac{1}{\sqrt{2}}  X^{-2} ( K \hook  *  \mathcal{F}  + \mathcal{F} \wedge \tilde{K} ) \, , \label{biY} \\
\dd ( X^{-1} \tilde{Y} ) & =&  \ii  ( X S ) H + \ii\tfrac{\sqrt{2}}{3}X^{-2}(K \hook  * B +B \wedge \tilde{K} ) \, , \label{bitY} \\
D ( X^{-1} Z ) & =& - \ii \sqrt{2}\, W  \, , \label{biO} \\
\dd V & =&  \sqrt{2} ( X + \tfrac{1}{3} X^{-3} ) *   Y   + \ii\tfrac{\sqrt{2}}{3}X^{-1}(\tilde{S}  * B + B\wedge Y)\nn\\
&&  - \tfrac{1}{\sqrt{2}}X^{-1}(S*\mathcal{F} + \mathcal{F}\wedge \tilde{Y})\, , \label{biV} \\
\dd \tilde{V} & =&  0 \, , \label{bitV} \\
D W & =&   - \tfrac{1}{\sqrt{2}}X^{-1} \mathcal{F} \wedge Z \, , \label{biW} \\
D \tilde{W} & =&  - \sqrt{2} ( X + \tfrac{1}{3} X^{-3} ) *Z  - \ii\tfrac{\sqrt{2}}{3}X^{-1} B \wedge Z \, ,\label{bitW}\\
\dd \Big[ ( X + \tfrac{1}{3} X^{-3} ) * Y \Big] & =&  \tfrac{\sqrt{2}}{3}\ii  B \wedge \tilde{V} - \tfrac{\ii}{3}X^{-1} H \wedge  Y  + \tfrac{1}{3\sqrt{2}}X^{-4} (* \mathcal{F}) \wedge  \tilde{K}  \,  \label{bistarY} .
\eea
}Here the covariant derivatives are $D=\diff + \ii\mathcal{A}\, \wedge$, and the contraction 
of a $p$-form $\rho$ into a $q$-form $\lambda$ (with $q\geq p$) is the $(q-p)$-form 
$(\rho\hook \lambda)_{\mu_1\cdots \mu_{q-p}}\equiv \frac{1}{p!} \rho^{\nu_1\cdots \nu_p}\lambda_{\nu_1\cdots \nu_p 
\mu_{1}\cdots\mu_{q-p}}$. 

In addition to  (\ref{biS}) -- (\ref{bistarY}) it is also straightforward to show that $K$ is a Killing one-form, so that the dual vector field 
$\xi\equiv K^{\#}$ is a Killing vector. We may hence introduce a local coordinate $\psi$, 
so that $\xi=\partial_\psi$ and the metric is independent of $\psi$. Since $K=S\sin 2\alphaangle\, K_1$, where $K_1$ has unit length, we may thus write
\bea
K_1 & = & S\sin 2\alphaangle\, (\diff \psi + \sigma)~, 
\eea
where $\mathcal{L}_\xi\sigma = 0 = i_\xi \sigma$ and $\mathcal{L}_\xi (S\sin 2\alphaangle)=0$. 
 
In order to analyse the equations (\ref{biS}) -- (\ref{bistarY}) further we write
\bea
B &=& B_1\wedge K_1 + B_\perp~, \qquad \mathcal{F} \ = \ \mathcal{F}_1\wedge K_1 + \mathcal{F}_\perp~,
\eea
where $B_1, B_\perp, \mathcal{F}_1, \mathcal{F}_\perp$ are chosen to have zero contraction with $K_1$.
The bilinear (\ref{biS}) then determines
\bea\label{B1}
B_1 &=& -\frac{3\ii}{\sqrt{2}S\sin 2\alphaangle}\diff (XS) -\ii X^{-2} K_2~.
\eea
Similarly the bilinear (\ref{bitS}) is equivalent to
\bea
\mathcal{F}_1 &=& -\frac{\sqrt{2}}{S\sin 2\alphaangle}\diff (XS\cos 2\alphaangle)~.
\eea
Contracting these last two equations with $K_1$, one concludes that 
$\mathcal{L}_\xi (XS) = 0 = \mathcal{L}_\xi \alphaangle$. Notice also that setting $\mathbb{A}=1_8$ in (\ref{DiracDilatino}) and taking the anti-commutator leads immediately 
to $\mathcal{L}_\xi X=0$. 
Having imposed (\ref{biS}), a short computation shows that 
equation (\ref{bitK}) is equivalent to  $\mathcal{L}_{\xi}B=0$.  One can also deduce 
from (\ref{bitK}) that $\mathcal{L}_{\xi}K_2=0$, and similarly from (\ref{bitS}) 
it follows that $\mathcal{L}_{\xi}\mathcal{F}=0$. We may then write
\bea\label{Agauge}
\mathcal{A} &=& -\sqrt{2}X\cot 2\alphaangle\, K_1 + \mathcal{A}_\perp~.
\eea
Notice  here we have made a partial gauge choice for $\mathcal{A}$.
Then
\bea\label{Fperp}
\mathcal{F}_\perp &=& -\sqrt{2}X S \cos 2\alphaangle\, \diff \sigma + \diff \mathcal{A}_\perp~.
\eea

Next one can show that equation (\ref{biK}) is equivalent to
\bea\label{Kbil}
X^2S^2\sin^2 2\alphaangle \, \diff \sigma &=& -\tfrac{2\sqrt{2}}{3}X^{-1}S\cos 2\alphaangle \,  J 
- \ii X^4 S\sin 2\alphaangle \, K_1\hook *H_\perp \nn\\
&&  + \sqrt{2} XS (\cos 2\alphaangle \, \mathcal{F}_\perp + \tfrac{2}{3}\ii B_\perp)~.
\eea
Here we have defined
\bea\label{Hperp}
H_\perp &\equiv & \ii \left[\tfrac{3}{\sqrt{2}}\diff (XS) + X^{-2} S\sin 2\alphaangle K_2\right]\wedge \diff \sigma + \diff B_\perp~.
\eea

The contractions of (\ref{biY}) and (\ref{bitY}) with $K_1$ imply that 
$\mathcal{L}_{\xi} J=0$. Equation (\ref{biY}) is then equivalent to
\bea\label{fatherofbarry}
\diff (X^{-1}SJ) &=& -\sqrt{2}S\sin 2\alphaangle\, J\wedge K_2 - \tfrac{3}{2\sqrt{2}}\cos 2\alphaangle \, \diff [(XS)^2\diff \sigma] + \ii XS \cos 2\alphaangle\, \diff B_\perp\nn\\&& 
- \tfrac{1}{\sqrt{2}}X^{-2} S \sin 2\alphaangle \left[K_1\hook *\mathcal{F}_\perp - K_2\wedge\mathcal{F}_\perp\right]~.
\eea
Similarly, one can show that (\ref{bitY}) is equivalent to
\bea\label{motherofbarry}
\diff (X^{-1}S\cos 2\alphaangle\, J) &=& -\tfrac{3}{2\sqrt{2}}  \diff[(XS)^2\diff\sigma] 
+ \ii XS \, \diff B_\perp \nn\\
&&+ \tfrac{\sqrt{2}}{3}\ii X^{-2}S\sin 2\alphaangle \left[K_1\hook *B_\perp 
- K_2\wedge B_\perp\right]~.
\eea
The contraction of equation (\ref{biO}) with $K_1$, in the gauge 
in which $\mathcal{A}$ is given by (\ref{Agauge}), 
 simply gives 
$\mathcal{L}_{\xi}\Omega=0$. 
Equation (\ref{biO}) is then equivalent to
\bea\label{Omegabil}
D_\perp (X^{-1}S\sin 2\alphaangle\, \Omega) &=& -\sqrt{2}S\Omega\wedge K_2~,
\eea
where $D_\perp \equiv \diff + \ii \mathcal{A}_\perp\wedge$. 

Finally we move onto the three-form bilinears. Equation (\ref{bitV}) states 
\bea
\diff (S\sin 2\alphaangle J\wedge K_2) &=& 0~.
\eea
The contraction of $K_1$ into (\ref{biW}) is equivalent to  (\ref{Omegabil}), while 
the remainder of this equation turns out to be the integrability condition for (\ref{Omegabil}). 
Next one can show that $K_1$  contracted into  (\ref{biV}) is implied by (\ref{fatherofbarry}) and (\ref{motherofbarry}),
while the remainder of this equation 
 reads
\bea\label{Jwedge}
-S^2\sin^2 2\alphaangle\,  J\wedge \diff\sigma &=& \sqrt{2}S\cos 2\alphaangle (X+\tfrac{2}{3}X^{-3})\tfrac{1}{2}J\wedge J - 2S K_1\hook *\diff \alphaangle \nn\\
&& -\tfrac{1}{\sqrt{2}}X^{-1}SJ\wedge (\cos 2\alphaangle\, \mathcal{F}_\perp + \tfrac{2}{3}\ii B_\perp)~.
\eea
Next we find that $K_1$ contracted into (\ref{bitW})
is implied by (\ref{Omegabil}). Using (\ref{Omegabil}) the remainder of this equation
reads
\bea\label{Omegawedge}
S^2\sin^2 2\alphaangle\, \Omega\wedge \diff\sigma \ =\  -2\ii S \diff \alphaangle\wedge K_2 
\wedge \Omega  + \tfrac{1}{\sqrt{2}}X^{-1}S\Omega\wedge (\cos 2\alphaangle\, 
\mathcal{F}_\perp + \tfrac{2}{3}\ii B_\perp)~.
\eea
The contraction of  $K_1$ into (\ref{bistarY}) can again be shown 
to follow from equations derived so far, while the remaining content of this 
equation is (on using various other equations) equivalent to
\begin{equation}
X^4 K_2 \hook \diff (X^{-3}S\sin 2\alphaangle) + \sqrt{2}S(X^2-\tfrac{2}{3}X^{-2}) + \tfrac{1}{\sqrt{2}}S J\hook
(\mathcal{F}_\perp + \tfrac{2}{3}\ii \cos 2\alphaangle B_\perp) \ = \ 0~.
\end{equation}

\section{More on the dilatino equation}\label{SecDil}

In the Abelian case of interest, the dilatino equation (\ref{dilatino}) may be written as $\delta \chi=0$, where we have introduced
\bea\label{dildef}
\delta \chi & \equiv & - \ii X^{-1} \partial_\mu X \Gamma^\mu \epsilon + \tfrac{1}{2\sqrt{2}} \left( X - X^{-3} \right) \Gamma_7 \epsilon + \tfrac{\ii}{24} X^2 H_{\mu\nu\rho} \Gamma^{\mu\nu\rho} \Gamma_7 \epsilon \nonumber \\
&&- \tfrac{1}{12\sqrt{2}} X^{-1} B_{\mu\nu} \Gamma^{\mu\nu} \epsilon - \tfrac{\ii}{8\sqrt{2}} X^{-1} \mathcal{F}_{\mu\nu} \Gamma^{\mu\nu} \Gamma_7 \epsilon~.
\eea
Recall here that $A_\mu^1=A_\mu^2=0$, while $\mathcal{A}_\mu\equiv A^3_\mu$, with curvature $\mathcal{F}=\diff \mathcal{A}$.
The right hand side of (\ref{dildef}) is an 8-component spinor, and thus $\delta\chi=0$ comprise 8 algebraic equations for $\epsilon=\epsilon_++\epsilon_-$.  

We begin by noting that 
neither of the definite chirality projections
$\epsilon_+$ nor $\epsilon_-$ can be identically zero. For 
if $\epsilon_\pm=0$, respectively, then we 
 in fact have an $SU(3)$ structure, rather than $SU(2)$ structure, and 
the bilinear $W\equiv \epsilon^\mt \Gamma_{(3)}\epsilon = \Omega_\mp$
 is the 
corresponding complex three-form.  However, since the left hand side of equation (\ref{biO}) 
of appendix~\ref{SecDiff} is identically zero in this case, we would deduce that $\Omega_\mp=0$ and 
hence $\epsilon_\mp=0$. 

On an open dense subset where $\epsilon_\pm$ 
are both non-zero, we then have that $\{\epsilon_\pm, \Gamma_\mu \epsilon_\pm^*\}$ 
span the positive and negative chirality spin bundles $\Spin^\pm$, respectively. 
Recall from (\ref{SU2rotation}) that $\epsilon_+=\sqrt{S}\cos \alphaangle \, \eta_1$, 
$\epsilon_- = \sqrt{S}\sin\alphaangle\, \eta_2^*$, where $\eta_1$ and $\eta_2$ have unit norm.
 In an orthonormal frame $(e^1,\ldots,e^4,e^5\equiv K_1,e^6\equiv K_2)$
in which the canonical $SU(2)$ structure defined by $\eta_1$ and $\eta_2$ is given by (\ref{SU2str}), 
one can easily check that $\{\epsilon_+,\Gamma_1\epsilon_+^*,\Gamma_3\epsilon_+^*,\Gamma_5\epsilon_+^*\}$ 
form a basis for  $\Spin^+$, while $\{\epsilon_-,\Gamma_1\epsilon_-^*,\Gamma_3\epsilon_-^*,\Gamma_5\epsilon_-^*\}$ 
form a basis for  $\Spin^-$. Thus
in order for the dilatino equation $\delta \chi=0$ to hold, it is sufficient to check that
the contraction of (\ref{dildef}) 
with $\epsilon_\pm^\dagger$ and $\epsilon_\pm^\mt\Gamma_\mu$ is zero. 
These are equivalent to two scalar and two one-form equations, respectively, that may be 
expressed in terms of the bilinears (\ref{bilinears}). Specifically, we may take the two
scalar contractions to be
\bea\label{dilscalar}
\epsilon^\dagger\delta \chi & = &- \ii X^{-1} \partial_\mu X K^\mu + \tfrac{1}{2\sqrt{2}} \left( X - X^{-3} \right) \tilde{S} + \tfrac{\ii}{24} X^2 H_{\mu\nu\rho} \tilde{V}^{\mu\nu\rho} \nonumber \\
&&+ \tfrac{\ii}{12\sqrt{2}} X^{-1} B_{\mu\nu} Y^{\mu\nu} - \tfrac{1}{8\sqrt{2}} X^{-1} \mathcal{F}_{\mu\nu} \tilde{Y}^{\mu\nu} \, ,\nn\\
\epsilon^\dagger \Gamma_7 \delta \chi & = &  X^{-1} \partial_\mu X \tilde{K}^\mu + \tfrac{1}{2\sqrt{2}} \left( X - X^{-3} \right) S - \tfrac{1}{24} X^2 H_{\mu\nu\rho} V^{\mu\nu\rho} \nonumber \\
&&+ \tfrac{\ii}{12\sqrt{2}} X^{-1} B_{\mu\nu} \tilde{Y}^{\mu\nu} - \tfrac{1}{8\sqrt{2}} X^{-1} \mathcal{F}_{\mu\nu} Y^{\mu\nu} \, ,
\eea
while the two one-form contractions are
\bea\label{diloneform}
\epsilon^\mathrm{T} \Gamma_\sigma \delta \chi & = & \tfrac{\ii}{8} X^2 H_{\mu\nu\sigma} Z^{\mu\nu} - \tfrac{1}{12\sqrt{2}} X^{-1} B^{\mu\nu} W_{\mu\nu\sigma} - \tfrac{\ii}{8\sqrt{2}} X^{-1} \mathcal{F}^{\mu\nu} \tilde{W}_{\mu\nu\sigma}~,\nn\\
\epsilon^\mathrm{T} \Gamma_\sigma \Gamma_7 \delta \chi & = & - \ii X^{-1} \partial^\mu X Z_{\mu\sigma} - \tfrac{1}{8} X^2 (*H)_{\mu\nu\sigma} Z^{\mu\nu} \nonumber \\
&&- \tfrac{1}{12\sqrt{2}} X^{-1} B^{\mu\nu} \tilde{W}_{\mu\nu\sigma} - \tfrac{\ii}{8\sqrt{2}} X^{-1} \mathcal{F}^{\mu\nu} W_{\mu\nu\sigma}~.
\eea
The dilatino equation $\delta \chi=0$ is thus equivalent to the the right hand sides of (\ref{dilscalar}) 
and (\ref{diloneform}) being zero. A tedious, but straightforward, calculation shows that
 $\delta\chi=0$ is implied by the differential 
constraints (\ref{diffconstraints}).

\section{Integrability conditions}\label{SecInt}

For what follows it will be convenient to record the component form of the Romans field equations in \eqref{FullEOM} and \eqref{Einstein}:
\bea
 \left( \mathscr{E}_g \right)_{\mu \nu} & \equiv &  R_{\mu \nu} - 4 X^{-2} \partial_\mu X \partial_\nu X - \left( \tfrac{1}{18} X^{-6} -\tfrac{1}{2} X^2 - \tfrac{2}{3} X^{-2} \right) g_{\mu\nu} \nn \\ & &- \tfrac{1}{4} X^4 ( H_\mu{}^{\rho \sigma} H_{\nu \rho \sigma} - \tfrac{1}{6} g_{\mu\nu} H^{\rho \sigma \tau} H_{\rho \sigma \tau} ) - \tfrac{2}{9} X^{-2} ( B_\mu{}^\rho B_{\nu\rho} - \tfrac{1}{8} g_{\mu\nu} B^{\rho\sigma} B_{\rho \sigma} )\nn \\ && - \tfrac{1}{2} X^{-2} ( F^{i\  \rho}_\mu{} F^i_{\nu \rho} - \tfrac{1}{8} g_{\mu\nu} F^{i\rho\sigma } F^i_{\rho\sigma} ) ~,\nn\\
\left( \mathscr{E}_X\right) &\equiv &  \nabla^\mu ( X^{-1} \partial_\mu X ) +  \left( \tfrac{1}{2} X^2 - \tfrac{2}{3} X^{-2} + \tfrac{1}{6} X^{-6} \right) - \tfrac{1}{24} X^4 H^{\mu\nu\rho} H_{\mu\nu\rho} \nn\\ &&+ \tfrac{1}{16} X^{-2} (\tfrac{4}{9} B^{\mu\nu} B_{\mu\nu} + F^{i\mu\nu} F^i_{\mu\nu} )~, \nn\\
\left( \mathscr{E}_A \right)^{\mu} &\equiv & \nabla_\nu ( X^{-2} B^{\nu\mu} ) - \tfrac{\ii}{12} \varepsilon^{\mu\nu\rho\sigma\tau\kappa} B_{\nu\rho} H_{\sigma\tau\kappa}~,\nn\\
\left( \mathscr{E}_{A^i} \right)^{\mu}& \equiv &  D_\nu ( X^{-2} F^{i\nu\mu } ) - \tfrac{\ii}{12}  \varepsilon^{\mu\nu\rho\sigma\tau\kappa} F^i_{\nu\rho}H_{\sigma\tau\kappa}~,\nn\\
 \left( \mathscr{E}_{B} \right)^{\mu \nu} & \equiv &  \nabla_\rho ( X^4 H^{\rho\mu\nu} ) - \tfrac{4}{9}  X^{-2} B^{\mu\nu} - \tfrac{\ii}{8} \varepsilon^{\mu\nu\rho\sigma\tau\kappa} (\tfrac{4}{9} B_{\rho\sigma} B_{\tau\kappa} + F^i_{\rho\sigma} F^i_{\tau\kappa} )~.
\eea
The equations of motion are then $\mathscr{E}_{\, \mathrm{field}}=0$. The field $A$ is the Stueckelberg 
one-form, that we set to zero  using the gauge symmetry of the theory. Its equation 
of motion 
$\mathscr{E}_A=0$ follows from taking the divergence of the $B$-field equation of motion
 $\mathscr{E}_B=0$.
We  also introduce
\begin{eqnarray}
 \left( \mathscr{B}_{F} \right)_{\mu\nu\rho}  &\equiv & \nabla_{[\mu} B_{\nu\rho]} - \frac{1}{3}  H_{\mu\nu\rho}~,\nn \\
 \left( \mathscr{B}_{F^i} \right)_{\mu\nu\rho}& \equiv &  D_{[\mu} F^i_{\nu\rho]}~,\nn \\ 
\left( \mathscr{B}_{H} \right)_{\mu\nu\rho\sigma} &\equiv & \nabla_{[ \mu} H_{\nu\rho\sigma ]}~.
\end{eqnarray}
Note that  $\mathscr{B}_{\, \mathrm{field}}$ vanish automatically as a consequence of the Bianchi identities.
For the Abelian case studied in the main text recall that $F_{\mu\nu}^1=F_{\mu\nu}^2=0$ while 
$ \mathcal{F}_{\mu\nu}\equiv F_{\mu\nu}^3 $.

In what follows we will show that 
supersymmetry together with $(\mathscr{E}_B)_\perp =0$  imply the equations of motion for all the fields. 
We begin by taking the exterior derivative of (\ref{biK}) to obtain
\begin{equation}
0 \ = \ - \tfrac{2\sqrt{2}}{3} \dd ( X^{-1} \tilde{Y} ) - \ii\, \dd (X^4 K \hook *H  ) + \tfrac{2\sqrt{2}}{3} \ii \, \dd [ X S B] - \sqrt{2} \mathcal{F} \wedge \dd ( X \tilde{S} ) \, .
\end{equation}
Using \eqref{biS}, \eqref{bitS} and  \eqref{bitY} then gives
\begin{equation}
0 \ = \ - \ii\, \dd (  X^4 K\hook *H ) ) - \tfrac{4}{9}\ii K \hook * B + \tfrac{4}{9} B \wedge (K\hook B) + \mathcal{F} \wedge (K\hook \mathcal{F}) \, .
\end{equation}
Since $\mathcal{L}_\xi ( X^4 * H )=0$ it hence follows that $K_1\hook \mathscr{E}_B=0$. 
Recall that
\bea
\mathscr{E}_B &=& K_1\wedge (K_1\hook \mathscr{E}_B) + (\mathscr{E}_B)_\perp~.
\eea
In general it is not true that supersymmetry implies $(\mathscr{E}_B)_\perp=0$. We henceforth 
impose this equation, and continue our analysis by 
taking the exterior derivative of (\ref{bistarY}). After a computation 
we find this implies
\bea\label{badger}
 \tfrac{2}{3}\ii \left[\diff(X^{-2}*B) + \ii B\wedge H\right](X\tilde{S}) - \left[\diff(X^{-2}*\mathcal{F}) + \ii \mathcal{F}\wedge H\right](XS) &=& 0~.
\eea
Since $\mathscr{E}_B=0$ implies $\mathscr{E}_A=0$, (\ref{badger}) implies $\mathscr{E}_{\mathcal{A}}=0$. 

To obtain the remaining equations of motion, we may use the integrability conditions 
for the dilatino equation \eqref{dilatino} and Killing spinor equation (\ref{KSE}) 
derived in \cite{Alday:2014bta}:
\bea
0
&=&  \ii \left(\mathscr{E}_X\right) \epsilon_I - \tfrac{1}{6\sqrt{2}} X \left( \mathscr{E}_A \right)_{\mu} \Gamma^\mu \epsilon_I - \tfrac{\ii}{4\sqrt{2}} X \left( \mathscr{E}_{A^i} \right)_{\mu} \Gamma^\mu \Gamma_7 ( \sigma_i )_I{}^J \epsilon_J + \tfrac{\ii}{8} X^{-2} \left( \mathscr{E}_{B} \right)_{\mu \nu} \Gamma^{\mu\nu} \Gamma_7 \epsilon_I \nn\\
&&- \tfrac{1}{12\sqrt{2}} X^{-1} \left( \mathscr{B}_{F} \right)_{\mu\nu\rho} \Gamma^{\mu\nu\rho} \epsilon_I - \tfrac{\ii}{8\sqrt{2}} X^{-1} \left( \mathscr{B}_{F^i} \right)_{\mu\nu\rho} \Gamma^{\mu\nu\rho} \Gamma_7 ( \sigma_i )_I{}^J \epsilon_J \nn\\&&+ \tfrac{\ii}{24} X^2 \left( \mathscr{B}_{H} \right)_{\mu\nu\rho\sigma} \Gamma^{\mu\nu\rho\sigma} \Gamma_7 \epsilon_I~,\label{dilin}\\
0
&=&  \tfrac{1}{2}\left(\mathscr{E}_X\right) \Gamma_\mu \epsilon_ I - \tfrac{1}{2} \left( \mathscr{E}_g \right)_{\mu \nu} \Gamma^\nu \epsilon_I  - \tfrac{1}{8} X^{-2} \left( \mathscr{E}_B \right)^{\nu\rho} \Gamma_{\mu\nu\rho} \Gamma_7 \epsilon_I \nn \\
&&-\tfrac{\ii}{3 \sqrt{2}} X \left( \mathscr{E}_A \right)_{\mu} \epsilon_I+\tfrac{1}{2 \sqrt{2}} X \left( \mathscr{E}_{A^{i}} \right)_{\mu} \Gamma_7 ( \sigma_{i} )_{I}{}^{J} \epsilon_J - \tfrac{1}{24} X^2 \left( \mathscr{B}_H \right)^{\nu \rho \sigma \tau} \Gamma_{\mu \nu \rho \sigma \tau} \Gamma_7 \epsilon_I\nn \\
&& - \tfrac{ \ii}{2 \sqrt{2}} X^{-1} \left( \mathscr{B}_F \right)_{\mu \nu \rho} \Gamma^{\nu \rho} \epsilon_I + \tfrac{3}{4 \sqrt{2}} X^{-1} \left( \mathscr{B}_{F^{i}} \right)_{\mu \nu \rho} \Gamma^{\nu \rho} \Gamma_7 ( \sigma_{i})_{I}{}^{J} \epsilon_J~.\label{KSEin}
\eea
Since $\mathscr{B}_{\, \mathrm{field}}=0$, and given the results above, 
(\ref{dilin}) immediately implies $\mathscr{E}_X=0$. Using this, 
and contracting (\ref{KSEin}) with $\epsilon^\dagger\Gamma^\upsilon$, we deduce 
the Einstein equation $\mathscr{E}_g=0$.

\section{Supersymmetry of the fundamental string}\label{SecCarolina}

In this appendix we show that the fundamental string considered in section 
\ref{SecWilson} is supersymmetric. 

As explained in \cite{Assel:2012nf} and \cite{Alday:2015lta}, BPS Wilson loops in the fundamental representation 
are dual to fundamental strings in the massive type IIA background $M_6\times S^4$. More precisely
 the string sits at the ``north pole'' of the four-sphere and wraps the $K_1$--$K_2$ direction 
 of the $SU(2)$ structure on $M_6$. Since the dual vector field to $K_1$ is proportional to the supersymmetric Killing 
 vector $\xi$, this means that the dual Wilson loop on the conformal boundary of $M_6$ wraps an orbit of $\xi$, 
 as expected from supersymmetry. It then remains to show that the fundamental string is itself 
 supersymmetric. This amounts to a certain projection  condition on the ten-dimensional 
 Killing spinor in massive IIA. Following a similar computation to  \cite{Assel:2012nf}, one 
 can show this reduces to the following projection condition on the six-dimensional 
 spinor $\epsilon$ on $M_6$:
\begin{equation}
 (1+\ii\Gamma_7\Gamma_{56})\epsilon \ = \ 0~. \label{proj6}
\end{equation}
Here recall that the orthonormal frame components are
$e^5=K_1$ and $e^6=K_2$. Recall also 
 from section \ref{SecSU2} that $\epsilon  =  \epsilon_+ + \epsilon_-$, 
where
\bea
\epsilon_+ &=& \sqrt{S}\cos\alphaangle\, \eta_1~, \qquad \epsilon_- \ = \ \sqrt{S}\sin\alphaangle\, \eta_2^*~.
\eea
The projection conditions  \cite{Gauntlett:2004zh} 
\begin{equation}
 \Gamma_7 \eta_1 \ = \  -\eta_1 \, , \qquad \Gamma_7 \eta_2^* \ = \  \eta_2^*~, \qquad 
 -\Gamma_{56} \eta_1 \ = \ \ii \eta_1 \, , \qquad -\Gamma_{56} \eta_2 \ = \ \ii \eta_2~,
\end{equation}
together with the fact that the Cliff$(6,0)$ matrices are purely imaginary then immediately 
imply that (\ref{proj6}) is indeed satisfied.
 Consequently the fundamental string wrapping the $K_1$--$K_2$ direction, at the north pole 
 of the internal $S^4$, is indeed 
 supersymmetric.



\begin{thebibliography}{}

\bibitem{Alday:2013lba} 
  L.~F.~Alday, D.~Martelli, P.~Richmond and J.~Sparks,
  ``Localization on Three-Manifolds,''
  JHEP {\bf 1310}, 095 (2013)
  [arXiv:1307.6848 [hep-th]].
  
\bibitem{Closset:2013vra} 
  C.~Closset, T.~T.~Dumitrescu, G.~Festuccia and Z.~Komargodski,
  ``The Geometry of Supersymmetric Partition Functions,''
  JHEP {\bf 1401}, 124 (2014)
  [arXiv:1309.5876 [hep-th]].
  
\bibitem{Imbimbo:2014pla}
  C.~Imbimbo and D.~Rosa,
  ``Topological anomalies for Seifert 3-manifolds,''
  JHEP {\bf 1507} (2015) 068
  [arXiv:1411.6635 [hep-th]].
  
\bibitem{Kallen:2012cs} 
  J.~K\"all\'en and M.~Zabzine,
  ``Twisted supersymmetric 5D Yang-Mills theory and contact geometry,''
  JHEP {\bf 1205}, 125 (2012)
  [arXiv:1202.1956 [hep-th]].
  
\bibitem{Hosomichi:2012ek} 
  K.~Hosomichi, R.~K.~Seong and S.~Terashima,
  ``Supersymmetric Gauge Theories on the Five-Sphere,''
  Nucl.\ Phys.\ B {\bf 865}, 376 (2012)
  [arXiv:1203.0371 [hep-th]].
  
  \bibitem{Kallen:2012va} 
  J.~K\"all\'en, J.~Qiu and M.~Zabzine,
 ``The perturbative partition function of supersymmetric 5D Yang-Mills theory with matter on the five-sphere,''
  JHEP {\bf 1208}, 157 (2012)
  [arXiv:1206.6008 [hep-th]].
 
\bibitem{Kim:2012ava} 
  H.~C.~Kim and S.~Kim,
  ``M5-branes from gauge theories on the 5-sphere,''
  JHEP {\bf 1305}, 144 (2013)
  [arXiv:1206.6339 [hep-th]].

\bibitem{Imamura:2012xg} 
  Y.~Imamura,
  ``Supersymmetric theories on squashed five-sphere,''
  PTEP {\bf 2013}, 013B04 (2013)
  [arXiv:1209.0561 [hep-th]].
  
\bibitem{Imamura:2012bm} 
  Y.~Imamura,
  ``Perturbative partition function for squashed $S^5$,''
  arXiv:1210.6308 [hep-th].

\bibitem{Qiu:2013pta} 
  J.~Qiu and M.~Zabzine,
  ``5D Super Yang-Mills on $Y^{p,q}$ Sasaki-Einstein manifolds,''
  Commun.\ Math.\ Phys.\  {\bf 333}, no. 2, 861 (2015)
  [arXiv:1307.3149 [hep-th]].

\bibitem{Qiu:2013aga} 
  J.~Qiu and M.~Zabzine,
  ``Factorization of 5D super Yang-Mills theory on $Y^{p,q}$ spaces,''
  Phys.\ Rev.\ D {\bf 89}, no. 6, 065040 (2014)
  [arXiv:1312.3475 [hep-th]].
  
\bibitem{Schmude:2014lfa} 
  J.~Schmude,
  ``Localisation on Sasaki-Einstein manifolds from holomorphic functions on the cone,''
  JHEP {\bf 1501}, 119 (2015)
  [arXiv:1401.3266 [hep-th]].
  
\bibitem{Pan:2013uoa} 
  Y.~Pan,
  ``Rigid Supersymmetry on 5-dimensional Riemannian Manifolds and Contact Geometry,''
  JHEP {\bf 1405}, 041 (2014)
  [arXiv:1308.1567 [hep-th]].

\bibitem{Imamura:2014ima} 
  Y.~Imamura and H.~Matsuno,
  ``Supersymmetric backgrounds from 5d $\mathcal N=$ 1 supergravity,''
  JHEP {\bf 1407}, 055 (2014)
  [arXiv:1404.0210 [hep-th]].

\bibitem{Pan:2014bwa} 
  Y.~Pan,
  ``5d Higgs Branch Localization, Seiberg-Witten Equations and Contact Geometry,''
  JHEP {\bf 1501}, 145 (2015)
  [arXiv:1406.5236 [hep-th]].
  
\bibitem{Alday:2015lta}
  L.~F.~Alday, P.~B.~Genolini, M.~Fluder, P.~Richmond and J.~Sparks,
  ``Supersymmetric gauge theories on five-manifolds,''
  JHEP {\bf 1508} (2015) 007
  [arXiv:1503.09090 [hep-th]].
   
\bibitem{Pan:2015nba}
  Y.~Pan and J.~Schmude,
  ``On rigid supersymmetry and notions of holomorphy in five dimensions,''
  JHEP {\bf 1511} (2015) 041
  [arXiv:1504.00321 [hep-th]].
  
\bibitem{Pini:2015xha}
  A.~Pini, D.~Rodriguez-Gomez and J.~Schmude,
  ``Rigid Supersymmetry from Conformal Supergravity in Five Dimensions,''
  JHEP {\bf 1509} (2015) 118
  [arXiv:1504.04340 [hep-th]].
  
\bibitem{Alday:2014rxa}
  L.~F.~Alday, M.~Fluder, P.~Richmond and J.~Sparks,
  ``Gravity Dual of Supersymmetric Gauge Theories on a Squashed Five-Sphere,''
  Phys.\ Rev.\ Lett.\  {\bf 113} (2014) 14,  141601
  [arXiv:1404.1925 [hep-th]].
  
\bibitem{Alday:2014bta}
  L.~F.~Alday, M.~Fluder, C.~M.~Gregory, P.~Richmond and J.~Sparks,
  ``Supersymmetric gauge theories on squashed five-spheres and their gravity duals,''
  JHEP {\bf 1409} (2014) 067
  [arXiv:1405.7194 [hep-th]].
  
\bibitem{Romans:1985tw}
  L.~J.~Romans,
  ``The F(4) Gauged Supergravity in Six-dimensions,''
  Nucl.\ Phys.\ B {\bf 269} (1986) 691.
  
\bibitem{Gauntlett:2004zh} 
  J.~P.~Gauntlett, D.~Martelli, J.~Sparks and D.~Waldram,
  ``Supersymmetric AdS(5) solutions of M theory,''
  Class.\ Quant.\ Grav.\  {\bf 21}, 4335 (2004)
  [hep-th/0402153].
    
\bibitem{Klare:2012gn} 
  C.~Klare, A.~Tomasiello and A.~Zaffaroni,
  ``Supersymmetry on Curved Spaces and Holography,''
  JHEP {\bf 1208}, 061 (2012)
  [arXiv:1205.1062 [hep-th]].
  
\bibitem{Klare:2013dka} 
  C.~Klare and A.~Zaffaroni,
  ``Extended Supersymmetry on Curved Spaces,''
  JHEP {\bf 1310}, 218 (2013)
  [arXiv:1308.1102 [hep-th]].

\bibitem{Hristov:2013spa} 
  K.~Hristov, A.~Tomasiello and A.~Zaffaroni,
  ``Supersymmetry on Three-dimensional Lorentzian Curved Spaces and Black Hole Holography,''
  JHEP {\bf 1305}, 057 (2013)
  [arXiv:1302.5228 [hep-th]].
     
\bibitem{Assel:2012nf}
  B.~Assel, J.~Estes and M.~Yamazaki,
  ``Wilson Loops in 5d N=1 SCFTs and AdS/CFT,''
  Annales Henri Poincare {\bf 15} (2014) 589
  [arXiv:1212.1202 [hep-th]].
  
\bibitem{Jafferis:2012iv} 
  D.~L.~Jafferis and S.~S.~Pufu,
  ``Exact results for five-dimensional superconformal field theories with gravity duals,''
  JHEP {\bf 1405}, 032 (2014)
  [arXiv:1207.4359 [hep-th]].
  
\bibitem{Cvetic:1999un}
  M.~Cvetic, H.~Lu and C.~N.~Pope,
  ``Gauged six-dimensional supergravity from massive type IIA,''
  Phys.\ Rev.\ Lett.\  {\bf 83} (1999) 5226
  [hep-th/9906221].
  
\bibitem{Qiu:2014oqa}
  J.~Qiu, L.~Tizzano, J.~Winding and M.~Zabzine,
  ``Gluing Nekrasov partition functions,''
  Commun.\ Math.\ Phys.\  {\bf 337} (2015) no.2,  785
  [arXiv:1403.2945 [hep-th]].
  
\bibitem{Ferrara:1998gv}
  S.~Ferrara, A.~Kehagias, H.~Partouche and A.~Zaffaroni,
  ``AdS(6) interpretation of 5-D superconformal field theories,''
  Phys.\ Lett.\ B {\bf 431} (1998) 57
  [hep-th/9804006].

\bibitem{Brandhuber:1999np}
  A.~Brandhuber and Y.~Oz,
  ``The D-4 - D-8 brane system and five-dimensional fixed points,''
  Phys.\ Lett.\ B {\bf 460} (1999) 307.
  
\end{thebibliography}
\end{document}